\documentclass[preprint,showpacs, showkeys,secnumarabic,amssymb,amsmath,amsfonts,aps,pra]{revtex4}

\usepackage{amsmath}
\usepackage{amsthm}

\begin{document}

\newcommand{\Zc}{\mbox{${\cal Z}_c$}}
\newcommand{\Nat}{{\cal I} \!\!\! {\cal N}}
\newcommand{\Ree}{{\Bbb R}}

\newcommand{\be}{\begin{eqnarray}}
\newcommand{\ee}{\end{eqnarray}}
\newcommand{\ba}{\begin{array}}
\newcommand{\ea}{\end{array}}
\newcommand{\bena}{\begin{eqnarray}}
\newcommand{\eena}{\end{eqnarray}}
\newcommand{\bdis}{\begin{displaymath}}
\newcommand{\edis}{\end{displaymath}}
\newcommand{\bit}{\begin{itemize}}
\newcommand{\eit}{\end{itemize}}
\newcommand{\ben}{\begin{enumerate}}
\newcommand{\een}{\end{enumerate}}
\newcommand{\nid}{\noindent}
\newcommand{\cl}{\centerline}
\newcommand{\nl}{\newline}
\newcommand{\ul}{\underline}
\newcommand{\dd}{\quad}
\newcommand{\re}{{\cal I \! \!R}}
\newcommand{\co}{\,{ I \! \! \! \! C}}
\newcommand{\ze}{{\cal Z \! \! \! \!Z}}
\newcommand{\zp}{{\cal Z}(\beta)}
\newcommand{\zpn}{{\cal Z}(\beta, N)}
\newcommand{\zpc}{{\cal Z }_c ( \beta )}
\newcommand{\zpcn}{{\cal Z }_c ( \beta,N )}
\newcommand{\isn}{\Omega(v,N)}
\newcommand{\isnp}{\Omega^\prime(v,N)}
\newcommand{\imp}{\int {\cal D }[q]}
\newcommand{\imf}{\int {\cal D }[\phi]}
\newcommand{\ebvq}{\exp \{ - \beta \sum_i  V[q_i] \}}
\newcommand{\ebvf}{\exp \{ - \beta \int dx   V[\phi(x)] \}}
\newcommand{\ebvfd}{\exp \{ - \beta \sum_i   V[\phi] \}}
\newcommand{\zee}{{\cal Z \! \! \! \!Z}_2}
\newcommand{\boh}{\Longleftrightarrow}
\newcommand{\void}{\lpg \o \rpg}
\newcommand{\mm}{{\rho_{\mu}}}
\newcommand{\mc}{{\rho_{can}}}
\newcommand{\mgc}{{\rho_{gc}}}
\newcommand{\ham}{{\cal H}}
\newcommand{\Xc}{{\cal X}}
\newcommand{\Yc}{{\cal Y}}
\newcommand{\Tc}{{\cal T}}
\newcommand{\Mc}{{\cal M}}
\newcommand{\Cc}{{\cal C}}
\newcommand{\vb}{\bar{v}}
\newcommand{\ub}{\bar{u}}
\newcommand{\vbx}{{v \! \! \! \!  -}_{max}}
\newcommand{\vbi}{{v \! \! \! \!  -}_{min}}
\newcommand{\Kc}{{\cal K}}
\newcommand{\Rc}{{\cal R}}
\newcommand{\Mcv}{{\cal M}_v}
\newcommand{\Mcg}{{\cal M}^G_g}
\newcommand{\Mce}{{\cal M}^{\cal H}_E}
\newcommand{\Si}{\Sigma}
\newcommand{\Siv}{\Sigma_v}
\newcommand{\Sig}{\Sigma^G_g}
\newcommand{\Sie}{\Sigma^{\cal H}_E}
\newcommand{\nG}{\frac{\partial^\mu G}{\| \nabla G \|}}
\newcommand{\nGn}{\frac{\partial^\mu G}{\| \nabla G \|^2}}
\newcommand{\nH}{\frac{\partial^\mu {\cal H}}{\| \nabla {\cal H} \|}}
\newcommand{\nHn}{\frac{\partial^\mu {\cal H}}{\| \nabla {\cal H} \|^2}}
\newcommand{\ngH}{\| \nabla {\cal H} \|}
\newcommand{\ngV}{\| \nabla V \|}
\newcommand{\gV}{ \nabla V}
\newcommand{\de}{\partial}
\newcommand{\too}{\longrightarrow}
\newcommand{\impl}{\Longrightarrow}
\newcommand{\eb}{{\bf e}}
\newcommand{\lpt}{\left(}
\newcommand{\rpt}{\right)}
\newcommand{\lpg}{\{}
\newcommand{\rpg}{\}}
\newcommand{\lpq}{\left[}
\newcommand{\rpq}{\right]}
\newcommand{\e}{\mbox{\boldmath $e$}}
\newcommand{\x}{\mbox{\boldmath $x$}}
\newcommand{\y}{\mbox{\boldmath $y$}}
\newcommand{\vi}{\mbox{\boldmath $i$}}
\newcommand{\vj}{\mbox{\boldmath $j$}}
\newcommand{\qs}{\frac{dq}{ds}}
\newcommand{\q}{\mbox{\boldmath $q$}}
\newcommand{\p}{\mbox{\boldmath $p$}}
\newcommand{\Q}{\mbox{\boldmath $Q$}}
\newcommand{\bP}{\mbox{\boldmath $P$}}
\newcommand{\bal}{\mbox{\boldmath $\alpha$}}
\newcommand{\A}{{\bf A}}
\newcommand{\bS}{{\bf S}}
\newcommand{\cn}{\mbox{cn}}
\newcommand{\sn}{\mbox{sn}}
\newcommand{\lps}{\langle}
\newcommand{\rps}{\rangle}
\newcommand{\dyy}{\displaystyle}
\newcommand{\nnv}{\frac{\nabla}{\| \nabla V \|}}
\newcommand{\unv}{\frac{1}{\| \nabla V \|}}
\newcommand{\ps}{\underline{\psi}}
\newcommand{\psv}{\underline{\psi}(V)}
\newcommand{\psc}{\underline{\psi}(\chi)}
\newcommand{\psvi}{{\psi_i}(V)}
\newcommand{\psvj}{{\psi_j}(V)}
\newcommand{\psvk}{{\psi_k}(V)}
\newcommand{\psvr}{{\psi_r}(V)}
\newcommand{\psci}{{\psi_i}(\chi)}
\newcommand{\pscj}{{\psi_j}(\chi)}
\newcommand{\psck}{{\psi_k}(\chi)}
\newcommand{\pscr}{{\psi_r}(\chi)}
\newcommand{\psvd}{\underline{\psi}(V) \cdot}
\newcommand{\pscd}{\underline{\psi}(\chi) \cdot}
\newcommand{\dpsv}{\underline{ \cdot \psi}(V)}
\newcommand{\dpsc}{\underline{ \cdot \psi}(\chi)}
\newcommand{\emme}{\nabla \frac{\nabla V}{\| \nabla V \|}}
\newcommand{\dvnv}{\frac{\triangle V}{\| \nabla V \|^2}}
\newcommand{\dv}{\triangle V}


\newtheorem{lemma}{Lemma}
\newtheorem{theorem}{Theorem}
\newtheorem{corollary}{Corollary}
\newtheorem{conjecture}{Conjecture}
\newtheorem{proposition}{Proposition}
\newtheorem{definition}{Definition}
\newtheorem{remark}{Remark}
\theoremstyle{definition}
\newtheorem*{pro}{Proof}

\hoffset 1truecm
\hsize 14truecm
\baselineskip=22pt


\title{Topology and Phase Transitions I. Preliminary Results}
\date{\today}

\author{Roberto Franzosi }

\affiliation{ Dipartimento di Fisica dell'Universit\`a di Firenze, Via G. Sansone 1,
I-50019 Sesto Fiorentino, and C.N.R.-I.N.F.M., Italy }

\author{Marco Pettini\footnote{Corresponding author. e-mail: pettini@arcetri.astro.it,
Phone: +39-055-2752282, Fax: +39-055-220039.} }

\affiliation{ Istituto Nazionale
di Astrofisica -- Osservatorio Astrofisico di Arcetri, Largo E. Fermi 5,
 50125 Firenze, Italy\\ and I.N.F.M., Unit\`a di Firenze, and I.N.F.N., Sezione di
Firenze}

\author{Lionel Spinelli }

\affiliation{ Centre de Physique Th\'eorique du C.N.R.S.,
Luminy Case 907, F-13288 Marseille Cedex 9, France }

\begin{abstract}
In this first paper, we demonstrate a theorem that establishes a first step
toward proving a {\it necessary} topological condition
for the occurrence of first or second order phase transitions: we prove
that the topology of certain submanifolds of configuration space
must necessarily change at the phase transition point.
The theorem applies to smooth, finite-range and confining
potentials $V$ bounded below, describing systems confined in finite regions of
space with continuously varying coordinates.
The relevant configuration space submanifolds are both the level sets
$\{ \Sigma_{v}:=V_N^{-1}(v)\}_{v \in{\Bbb R}}$ of the
potential function $V_N$ and the configuration space submanifolds enclosed by
the $ \Sigma_{v}$ defined by
$\{ M_{v}:=V_N^{-1}((-\infty,v])\}_{v \in{\Bbb R}}$,
which are labeled by the potential energy value $v$, and where
$N$ is the number of degrees of freedom.
The proof of the  theorem proceeds by showing
that, under the assumption of diffeomorphicity of the equipotential
hypersurfaces $\{ \Sigma_{v}\}_{v\in{\Bbb R}}$, as well as of the
$\{ M_{v}\}_{v\in{\Bbb R}}$, in an arbitrary interval
of values for $\vb=v/N$, the Helmoltz free
energy is uniformly convergent in $N$ to its thermodynamic limit, at least
within the class of twice differentiable functions, in the corresponding
interval of temperature. This preliminary theorem is essential to prove another
theorem - in paper II -
which makes a stronger statement about the relevance of topology for phase
transitions.
\end{abstract}

 \maketitle

\section{Introduction}
\label{introduction}
Some years ago, based on the well known fact that an Hamiltonian flow corresponds
to a geodesic flow on a suitably defined Riemannian manifold, a new explanation of
the origin of Hamiltonian chaos has been proposed\cite{Pettini,pre96}.
With the aid of this "geometric
viewpoint", the dynamical and geometrical signatures of phase transitions have been
investigated in several models\cite{cccp,pre98,jpa98,CSCP,Firpo}.
Invariably, the occurrence of a
phase transition is signaled by a "cuspy" pattern of some curvature property of the
underlying mechanical Riemannian manifold, whereas no particular pattern is displayed
in the absence of a phase transition.
On the basis of an heuristic argument\cite{cccp,pre98}, it
has been conjectured that the observed geometric signatures of phase transitions could
be the consequence of a change of the topology of the mechanical manifolds.
After intermediate steps\cite{top1,top3}, direct evidence has been given of the actual
existence of
topological signatures of phase transitions. These have been put in evidence through the
numerical computation of the Euler characteristic (a topologic invariant) for the
level sets $\{ \Sigma_v\}_{v\in{\Bbb R}}$ of the potential function of a
two-dimensional lattice $\varphi^4$
model\cite{top2}, through the exact analytic computation
of the Euler characteristic of $\{ M_v = V_N^{-1}((-\infty,v])\}_{v\in{\Bbb R}}$
submanifolds of configuration space for a mean-field $XY$ model\cite{xymf} and for
a $k$-trigonometric model\cite{ptrig}.

These results have motivated the effort to make a leap forward by proving that topology
changes of configuration space submanifolds (either $\Sigma_v$ or $M_v$) are necessary
for the occurrence of phase transitions, at least for a class of potentials of physical
relevance.

In the present paper, a result of this kind is actually proved in the form of a necessity
theorem. However, one of its basic hypotheses is somewhat too restrictive -- and cannot
be relaxed in the present demonstration scheme -- to directly use our Main Theorem as
an evident rigorous support of our former {\it topological hypothesis}\cite{physrep}.
This notwithstanding, the Main Theorem proved in the present paper is indispensable to
prove a definitely stronger result, of a broad domain of applicability, given in paper II.

In the present paper, we prove the following theorem:

\smallskip
\noindent{\bf Theorem\ 1.} {\it Let $V_N(q_1,\dots,q_N):
{\Bbb R}^N \rightarrow{\Bbb R}$, be a smooth, non-singular, finite-range
potential. Denote by $\Sigma_v:= V_N^{-1}(v)$, $v\in{\Bbb R}$, its
{\em level sets}, or {\em equipotential hypersurfaces}, in
configuration space.

Then let $\vb =v/N$ be the potential energy per degree of freedom.

If for any pair of values $\vb$ and $\vb^\prime$ belonging  to a given
interval $I_{\vb}=[\vb_0, \vb_1]$ and for any $N>N_0$ it is

\centerline{$\Sigma_{N\vb}\approx \Sigma_{N\vb^\prime}$ }
\vskip 0.2truecm
that is $\Sigma_{N\vb}$ is {\em diffeomorphic} to $\Sigma_{N\vb^\prime}$,
then the sequence of the Helmoltz free energies
$\{ F_N (\beta)\}_{N\in{\Bbb N}}$ -- where $\beta =1/T$ ($T$ is the
temperature) and
$\beta\in I_\beta =(\beta (\vb_0), \beta (\vb_1))$ -- is {\em uniformly}
convergent at least in ${\cal C}^2(I_\beta)$ so that
$F_\infty \in{\cal C}^2(I_\beta)$ and neither first nor second order phase
transitions can occur in the (inverse) temperature interval
$(\beta (\vb_0), \beta (\vb_1))$. }
\smallskip

This is our first Theorem, given in Section \ref{mainthm}. Now, for any given
model
described by a smooth, non-singular, finite-range potential, it is in general
a hard task to locate all its critical points and thus
to ascertain whether the theorem actually applies to it or not. Therefore we
use Theorem 1 to prove - in paper II - a second theorem which, making a direct
link between thermodynamic entropy and a weighed sum of the Morse indexes of
the  submanifolds $M_v$, provides a general and stronger
result about the relevance of configuration space topology for phase
transitions. We anticipate below the formulation of this second theorem:

\smallskip
\noindent{\bf Theorem\ 2.} {\it Let $V_N(q_1,\dots,q_N): {\Bbb R}^N
\rightarrow {\Bbb R}$, be a smooth, non-singular, finite-range
potential. Denote by $M_v:= V_N^{-1}((-\infty,v])$, $v\in{\Bbb R}$, the
generic submanifold of configuration space bounded by $\Sigma_v$.
Let $\{ q_c^{(i)}\in{\Bbb R}^N\}_{i\in[1,{\cal N}(v)]}$ be the set of critical
points of the potential, that is s.t. $\nabla V_N(q_c^{(i)})=0$, and
${\cal N}(v)$ be the number of critical points up to the potential energy
value $v$. Let $\Gamma(q_c^{(i)},\varepsilon_0)$ be pseudo-cylindrical
neighborhoods of the critical points, and $\mu_i(M_v)$ be the Morse indexes
of $M_v$, then there exist real numbers $A(N,i,\varepsilon_0)$, $g_i$
and real smooth functions
$B(N,i,v,\varepsilon_0)$ such that the following equation for the
microcanonical configurational entropy $S_N^{(-)}(v)$ holds

\begin{eqnarray}
S_N^{(-)}(v) &=&\frac{1}{N} \log \left[ \int_{M_v
\setminus\bigcup_{i=1}^{{\cal N}(v)}
\Gamma(q^{(i)}_c,\varepsilon_0)}\ d^Nq + \sum_{i=0}^N
A(N,i,\varepsilon_0 ) \ g_i\ \mu_i (M_{v-\varepsilon_0})\right. \nonumber\\
&+&\left. \sum_{n=1}^{{\cal
N}_{cp}^{\nu(v)+1}}B(N,i(n),v-v_c^{\nu(v)},\varepsilon_0 )
  \right] \ ,\nonumber
\end{eqnarray}
(details and definitions are given in Section $2$ of paper II),
and an unbound growth with $N$ of one of the derivatives
$\vert\partial^k S^{(-)}(v)/\partial v^k\vert$, for $k=3,4$, and thus the
occurrence of a first or of a second order phase transition respectively, can
be entailed only by the topological term $\sum_{i=0}^N A(N,i,\varepsilon_0 )\
g_i\ \mu_i (M_{v-\varepsilon_0})+ \sum_{n=1}^{{\cal
N}_{cp}^{\nu(v)+1}}B(N,i(n),v-v_c^{\nu(v)},\varepsilon_0 )$. }
\smallskip

\noindent Together, these two theorems imply that for a wide class of
potentials which are good Morse functions, a first or a second order phase
transition can only be the consequence of a topology change of the submanifolds
$M_v$ of configuration space.

The converse is not true: topology changes are necessary but not sufficient for
the occurrence of phase transitions.
As we point out in Remark \ref{sufficiency},
the above mentioned works in Refs.\cite{top2} and \cite{xymf,ptrig} provide
some hints
about the sufficiency conditions but rigorous results are not yet available.

The reader can get a hold of the meaning of the main result of the present paper by
reading just Section \ref{sec2}, Section \ref{mainthm} and the beginning of
Section \ref{proofLm4} where a sketch of the proof of Lemma \ref{derivees-majorees} is
given. In Section \ref{mainthm} we enunciate the Main Theorem, four main Lemmas (and give the
short proofs of two of them), we give the condensed proof of the Main Theorem, we enunciate
a Corollary to the Main Theorem and give its proof.

Section \ref{proofLm4}, apart from the already mentioned sketch of the proof of
Lemma \ref{derivees-majorees}, which is the core of the proof of
Theorem 1, contains the most tedious and hard reading part of the paper which is
necessary to prove the Main Theorem but not to understand the meaning of the Theorem
itself.

A preliminary account of Theorem 1 has been given in Ref. \cite{pirl}.


\section{Basic definitions}
\label{sec2}

For a physical system ${\cal S}$ of $n$ particles confined in a
bounded subset  $\Lambda^d$ of ${\Bbb R}^d$, $d=1,2,3$,  and
interacting through a real valued
potential function $V_N$ defined on
$(\Lambda^d)^{\times n}$, with $N=n d$, the {\it configurational
microcanonical volume}
$\Omega(v,N)$ is defined for any value $v$ of the potential $V_N$ as
\begin{equation}
\Omega(v,N) = \int_{(\Lambda^d)^{\times n}} dq_1\dots dq_N\
\delta[V_N(q_1,\dots, q_N) - v]
= \int_{\Sigma_v}\ \frac{d\sigma}{\Vert \nabla V_N\Vert}~,
\label{mi_volume}
\end{equation}
where $d\sigma$ is a surface element of $\Si_v:=V_N^{-1}(v)$; in what
follows $\Omega(v,N)$ is also called {\it structure integral}. The norm
$\Vert \nabla V_N\Vert$ is defined as $\Vert \nabla V_N\Vert =[\sum_{i=1}^N
(\partial_{q_i}V_N)^2]^{1/2}$.
The {\it configurational partition function} $Z_c(\beta, N)$ is
defined as
\begin{equation}
Z_c(\beta, N) = \int_{(\Lambda^d)^{\times n}}dq_1\dots dq_N\
\exp [-\beta V_N(q_1,\dots, q_N)]
= \int_0^\infty dv\ e^{-\beta v}\int_{\Sigma_v}\ \frac{d\sigma}
{\Vert \nabla V_N\Vert}~,
\label{Zconf}
\end{equation}
where the real parameter $\beta$ has the physical meaning of an
inverse temperature. Notice that the formal Laplace transform of the
structure integral in the r.h.s. of  (\ref{Zconf}) stems from a co-area
formula \cite{federer} which is of very general validity (it holds also for
Hausdorff measurable sets).

Now we can define the configurational thermodynamic functions to be used in
this paper.

\begin{definition}
Using the notation  $\vb = v /N$ for the value of the potential energy per
particle, we introduce the following functions:

- {\em Configurational microcanonical entropy, relative
  to $\Sigma_v$}.
  For any $N\in{\Bbb N}$ and
       $\vb\in{\Bbb R}$,
    \begin{eqnarray}
     S_N(\vb)\equiv S_N(\vb;V_N)
          =\frac{1}{N}
      \log{\Omega(N \vb, N)} \, .
    \nonumber
    \end{eqnarray}

- {\em Configurational canonical free energy}.
   For any $N\in{\Bbb N}$ and $\beta\in{\Bbb R}$,
    \bena
    f_N(\beta)\equiv f_N(\beta; V_N)=
        \frac{1}{N} \log Z_c(\beta, N)\, . \nonumber
    \eena

- {\em Configurational microcanonical entropy,
        relative to the volume bounded by $\Sigma_v$}. For any $N\in{\Bbb N}$ and
        $\vb\in{\Bbb R}$,
        \begin{eqnarray}
      S^{(-)}_N(\vb) \equiv S^{(-)}_N(\vb;V_N)
          =\frac{1}{N} \log{ M (N \vb, N)} \,
    \nonumber
    \end{eqnarray}
where
\begin{equation}
M (v,N) = \int_{(\Lambda^d)^{\times n}} dq_1\dots dq_N\
\Theta [V_N(q_1,\dots, q_N) - v]
=\int_0^v d\eta \  \int_{\Sigma_\eta}\ \frac{d\sigma}{\Vert \nabla V_N\Vert}~,
\label{pallaM}
\end{equation}
with $\Theta[\cdot]$ the Heaviside step function; $M(v,N)$ is the
codimension-0 subset of configuration space enclosed by the equipotential
hypersurface $\Sigma_v$. The representation of $M(v,N)$ given in the r.h.s.
stems from the already mentioned co-area formula in \cite{federer}.
Moreover,  $S^{(-)}_N(\vb)$ is related with the configurational canonical
free energy, $f_N$, for any $N\in{\Bbb N}$ and $\vb\in{\Bbb R}$, through
the Legendre transform \cite{ruelle}
    \begin{eqnarray}
     - f_N(\beta) = \inf_{\vb} \{ \beta \cdot \vb - S^{(-)}_N(\vb)\}
    \, ,
    \end{eqnarray}
yielding, for any $N\in{\Bbb N}$ and $\beta \in{\Bbb R}$,
    \begin{eqnarray}
     - f_N(\beta) =  \beta \cdot \vb_N -  S^{(-)}_N(\vb_N)
    \label{legendre-tras}
    \eena
with, for any $N\in{\Bbb N}$ and $\vb\in{\Bbb R}$,
    \begin{eqnarray}
    \beta_N(\vb)=
    \frac{\partial S_N^{(-)}}{\partial \vb} (\vb) \, ,
    \label{vb2beta}
    \end{eqnarray}
and the inverse relation, valid for any $N\in{\Bbb N}$ and $\beta\in{\Bbb R}$,
    \begin{eqnarray}
    \vb_N(\beta)=-\frac{\partial f_N}{\partial \beta} (\beta)
    \, .
    \label{beta2vb}
    \end{eqnarray}
\label{def-elcc}
\end{definition}
Finally, for a system described by a Hamiltonian function $H$
of the kind $H=\sum_{i=1}^N p_i^2/2 + V_N(q_1,\dots,q_N)$,
the Helmoltz free energy is defined by
\begin{equation}
F_N(\beta; H)=-(N\beta)^{-1}\log
\int d^Np\ d^Nq\ \exp[-\beta H(p,q)]~,
\end{equation}
whence
\begin{equation}
F_N(\beta; H)=-(2\beta)^{-1}\log(\pi/\beta)-f_N(\beta, V_N)/\beta
\end{equation}
with
its thermodynamic limit ($N\rightarrow\infty$ and
${ vol}(\Lambda^d)/N={ const}$)
\begin{equation}
F_\infty(\beta)=\lim_{N\rightarrow\infty}F_N(\beta ; H)~.
\end{equation}

\begin{definition}[First and second order phase transitions]
\label{PTs}
We say that a physical system ${\cal S}$ undergoes a phase transition if
there exists a thermodynamic function which -- in the thermodynamic limit
($N\rightarrow\infty$ and ${ vol}(\Lambda^d )/N={ const}$) -- is
only piecewise analytic.
In particular, if the
first-order derivative of the Helmoltz free energy $F_\infty(\beta)$ is
discontinuous at some point $\beta_c$, then we say that a {\it first-order}
phase transition occurs.
If the second-order derivative of the Helmoltz free energy
$F_\infty(\beta)$ is
discontinuous at some point $\beta_c$, then we say that a {\it second-order}
phase transition occurs.
\end{definition}

\begin{definition}[Standard potential, fluid case]
\label{st-pot}
We say that an $N$ degrees of freedom potential $V_N$ is a
{\emph standard potential} for a fluid if it is of the form
\begin{eqnarray}
        V_N:& & {\cal B}_N\subset{\Bbb R}^N \rightarrow {\Bbb R}\nonumber\\
    V_N(q) &=& \sum_{i\neq j=1}^n
    \Psi (\| \vec{q}_i - \vec{q}_j  \|)
    + \sum_{i =1}^n U_\Lambda ( \vec{q}_i )\,
\end{eqnarray}
where ${\cal B}_N$ is a compact subset of ${\Bbb R}^N$, $N=nd$,
$\Psi$ is a real valued function of one variable such that additivity
holds, and where $U_\Lambda$ is any smoothed potential barrier to confine the
particles in a finite volume $\Lambda$, that is
\[
U_\Lambda ( \vec{q} )=\left\{ \begin{array}{cc}
0&~if~\vec{q}\in\Lambda^\prime \\
+\infty&~if~\vec{q}\in\Lambda^c,~ complement~ in~ {\Bbb R}^N \\
{\cal C}^\infty &function~for~\vec{q}\in\Lambda\setminus\Lambda^\prime
\end{array} \right.
\]
where $\Lambda^\prime\subset\Lambda$ and $\Lambda^\prime$ arbitrarily close
to $\Lambda\subset{\Bbb R}^N$, closed and bounded. $U_\Lambda$ is a confining
potential in a limited spatial volume with the additional property that
given two limited $d$-dimensional regions of space, $\Lambda_1$ and
$\Lambda_2$, having in common a $d-1$-dimensional boundary,
$U_{\Lambda_1}+U_{\Lambda_2}=U_{\Lambda_1\cup\Lambda_2}$.
By additivity we mean what follows. Consider two systems ${\cal S}_1$ and
${\cal S}_2$, having $N_1=n_1 d$ and $N_2=n_2 d$ degrees of freedom, occuping
volumes
$\Lambda_1^d$ and $\Lambda_2^d$, having potential energies $v_1$  and $v_2$,
for any $(q_1, \ldots ,q_{N_1})\in (\Lambda_1^d)^{\times n_1}$ such that
$V_{N_1}(q_1, \ldots ,q_{N_1}) =v_1$,
for any $(q_{N_1+1}, \ldots ,q_{N_1+N_2})\in (\Lambda_2^d)^{\times n_2}$
such that $V_{N_2}(q_{N_1+1},\ldots ,q_{N_1+N_2}) = v_2$,
for  $(q_1, \ldots ,q_{N_1+N_2})\in (\Lambda_1^d)^{\times n_1}\times
(\Lambda_2^d)^{\times n_2}$ let $V_N(q_1,\ldots ,q_{N_1+N_2}) = v$ be
the potential energy $v$ of the compound system
${\cal S}={\cal S}_1+{\cal S}_2$
which occupies the volume  $\Lambda^d = \Lambda_1^d\cup\Lambda_2^d$ and
contains $N=N_1+N_2$ degrees of freedom. If
    \begin{equation}
         v(N_1+N_2,\Lambda_1^d\cup\Lambda_2^d) = v_1(N_1,\Lambda_1^d)
      + v_2(N_2,\Lambda_2^d) + v^\prime (N_1,N_2,\Lambda_1^d,\Lambda_2^d)
        \label{V_somma}
    \end{equation}
where $v^\prime$ stands for the interaction energy between ${\cal S}_1$ and
${\cal S}_2$, and if $v^\prime /v_1\rightarrow 0$ and
$v^\prime /v_2\rightarrow 0$ for $N\rightarrow\infty$ then $V_N$ is additive.
Moreover, at short distances $\Psi$ must be a repulsive
potential so as to prevent the concentration of an arbitrary number of particles
within small, finite volumes of any given size.
\end{definition}

\begin{definition}[Standard potential, lattice case]
\label{st-pot1}
We say that an $N$ degrees of freedom potential $V_N$ is a
{\emph standard potential} for a lattice if it is of the form
\begin{eqnarray}
        V_N:& & {\cal B}_N\subset{\Bbb R}^N \rightarrow {\Bbb R}\nonumber\\
    V_N(q) &=& \sum_{{\underline i},{\underline j}\in{\cal I}\subset{\Bbb N}^d}
 C_{{\underline i} {\underline j}}
    \Psi (\| \vec{q}_{\underline i} - \vec{q}_{\underline j}  \|) +
    \sum_{{\underline i}\in{\cal I}\subset{\Bbb N}^d}
     \Phi ( \vec{q}_{\underline i} )
\end{eqnarray}
where ${\cal B}_N$ is a compact subset of ${\Bbb R}^N$. Denoting by
$a_1,\dots,a_d$ the lattice spacings, if ${\underline i}\in{\Bbb N}^d$, then
$(i_1a_1,\dots,i_da_d)\in\Lambda^d$. We denote by $m$ the number of lattice
sites in each spatial direction, by $n=m^d$ the total number of lattice sites,
by $D$ the number of degrees of freedom on each site. Thus
$\vec{q}_{\underline i}\in{\Bbb R}^D$ for any ${\underline i}$. The total
number of degrees of freedom is $N=m^dD$. Having two systems made of $N=m^dD$
degrees of freedom, whose site indexes
${ i}^{(1)}$ and ${ i}^{(2)}$ run over
$1\leq {i}_1^{(1)},\dots, { i}_d^{(1)}\leq m$, and
$1\leq {i}_1^{(2)},\dots, { i}_d^{(2)}\leq m$, after
gluing together the two systems through a common $d-1$ dimensional boundary
the new system has indexes ${ i}$ running over, for example,
$1\leq {i}_1\leq 2m$ and $1\leq {i}_2,\dots,
{i}_d\leq m$. If
\begin{equation}
        v(N+N,\Lambda_1^d\cup\Lambda_2^d) = v_1(N,\Lambda_1^d)
   + v_2(N,\Lambda_2^d) + v^\prime (N,N,\Lambda_1^d,\Lambda_2^d)
 \label{V_summa}
 \end{equation}
where $v^\prime$ stands for the interaction energy between the two systems and
if $v^\prime /v_1\rightarrow 0$ and
$v^\prime /v_2\rightarrow 0$ for $N\rightarrow\infty$ then $V_N$ is additive.
\end{definition}

\begin{definition}[Short-range potential]
\label{sr-pot}
In defining a short-range potential, a distinction has to be made between
lattice systems and fluid systems.
Given a standard potential $V_N$ on a lattice, we say that it is a short-range
potential if the coefficients $C_{{\underline i}{\underline j}}$
are such that for any ${\underline i},{\underline j}\in{\cal I}\subset{\Bbb N}^d$,
$C_{{\underline i}{\underline j}}=0$
iff $\vert {\underline i} - {\underline j}\vert > c$, with $c$ is
definitively constant for
$N\rightarrow\infty$.

Given a standard potential $V_N$ for a fluid system, we say that it is a
short-range potential if there exist $R_0>0$ and $\epsilon >0$ such that
for $\Vert{\bf q}\Vert >R_0$ it is
$\vert\Psi (\Vert{\bf q}\Vert )\vert < \Vert{\bf q}\Vert^{-(d+\epsilon)}$,
where $d=1,2,3$ is the spatial dimension.
\end{definition}

\begin{definition}[Stable potential]
\label{stab-pot}
We say that a potential $V_N$ is stable \cite{ruelle} if there exists
$B\geq 0$ such that
\begin{equation}
V_N(q_1,\dots,q_N)\geq -N B
\end{equation}
for any $N > 0$ and $(q_1,\dots,q_N)\in (\Lambda^d)^{\times n}$, or for
$\vec{q}_{\underline i}\in{\Bbb R}^D$,
${\underline i}\in{\cal I}\subset{\Bbb N}^d$, $N=m^dD$, for lattices.
\end{definition}

\begin{definition}[Confining potential]
\label{conf-pot}
With the above definitions of standard potentials $V_N$, in the fluid case the
potential is said to be confining in the sense that it contains $U_\Lambda$
which constrains the
particles in a finite spatial volume, and in the lattice case the potential
$V_N$ contains an on-site potential such that -- at finite energy --
$\Vert \vec{q}_{\underline i}\Vert$ is constrained in compact set of values.
\end{definition}

\begin{remark}[Compactness of equipotential hypersurfaces]
From the previous definition it follows that, for a confining potential, the
equipotential hypersurfaces $\Sigma_v$ are compact (because they are closed
by definition and bounded in view of particle confinement).
\end{remark}
\begin{proposition}[Pointwise convergence]
\label{pro-elcc}
Assume $V_N$ is a standard, confining, short-range and stable potential.
Assume also that there exists $N_0\in{\Bbb N}$ such that
$\bigcap_{N>N_0}^\infty{ dom}(S^{(-)}_N)$ and
$\bigcap_{N>N_0}^\infty{dom}(S_N)$ are nonempty sets,
then the following pointwise limits exist almost everywhere
\begin{eqnarray}
& &\lim_{N \too \infty} S^{(-)}_N(\vb) \equiv S_\infty^{(-)}(\vb)~~~
{\rm for}~~~
\vb\in\bigcap_{N>N_0}^\infty{dom}(S^{(-)}_N)
\nonumber \\
& &\lim_{N \too \infty} S_N(\vb) \equiv S_\infty(\vb)~~~{\rm for}~~~
\vb\in\bigcap_{N>N_0}^\infty{ dom}(S_N)
\nonumber
\end{eqnarray}
and moreover
\begin{equation}
S_\infty^{(-)} (\vb) = S_\infty (\vb)~~~{\rm for}~~~
\vb\in \bigcap_{N>N_0}^\infty{dom}(S^{(-)}_N)\cap
\bigcap_{N>N_0}^\infty{ dom}(S_N)
\nonumber
\end{equation}
\end{proposition}

\begin{pro}
\label{pointconv}
The existence of the thermodynamic limit for the sequences of functions
$S^{(-)}_N$ and $S_N$, associated with a standard potential function $V_N$
with short-range interactions, stable and confining is formally proved in
\cite{ruelle}, chapters 3.3 and 3.4.
To prove that in the thermodynamic limit the two entropies  $S_\infty^{(-)}$
and
$S_\infty$ are equal, we proceed from the definitions of $S^{(-)}_N$ and of
$\beta_N(\vb)$, that is
\[
 S^{(-)}_N(\vb)=\frac{1}{N} \log{ M (N \vb, N)} \,
\]
and
\[
\beta_N(\vb)=
\frac{\partial S_N^{(-)}}{\partial \vb} (\vb) \, ,
\]
noting that from the r.h.s. of Eq.(\ref{pallaM}) we obtain
\begin{equation}
\frac{d M (N \vb, N)}{d \vb} = N \Omega (N \vb, N)
\label{derivemme}
\end{equation}
so that
\begin{equation}
\beta_N(\vb)=\frac{1}{N M (N \vb, N)}\frac{d M (N \vb, N)}{d \vb}=
\frac{\Omega (N \vb, N)}{M (N \vb, N)}
\label{betaenne}
\end{equation}
whence
\begin{equation}
\frac{1}{N}\log \Omega(\vb N,N) = \frac{1}{N}\log M (\vb N,N) +
\frac{1}{N}\log \beta_N(\vb)\ .
\label{equi}
\end{equation}
Because of the existence of the thermodynamic limit $\beta(\vb)$ of the
sequence of functions
$\beta_N(\vb)$ [see Proposition 2], for any given $\vb\in{\Bbb R}$ it is
\[
\lim_{N\to\infty}\frac{1}{N}\log \beta_N(\vb) = 0
\]
thus, being $S_N(\vb)=1/N\log \Omega(\vb N,N)$, in the thermodynamic limit,
that is in the limit $N\to\infty$ with $vol(\Lambda^d)/N= const$,
for any $\vb\in{\Bbb R}$ Eq.(\ref{equi}) implies
    \bena
    S_\infty (\vb) = S_\infty^{(-)} (\vb)~.
    \label{finf-ginf}
    \eena
\qed
\end{pro}
\begin{remark}[Equivalent definitions of entropy]
In Ref.\cite{ruelle} it is proved that the Legendre transform relating
$S^{(-)}_N (\vb)$ with $f_N(\beta)$ still holds true in the thermodynamic
limit, that is $S_\infty^{(-)} (\vb)$ and $f_\infty(\beta)$ are still
related by
a Legendre transform (see theorem 3.4.4 at p.55 of Ref.\cite{ruelle}). Thus,
after equation (\ref{finf-ginf}) also $S (\vb)$ is related with
$f_\infty(\beta)$ by the same Legendre transform.
\end{remark}

\begin{proposition}[Pointwise convergence]
\label{pro-conv}
Assume $V_N$ is a standard, confining, short-range and stable potential.
Assume also that there exists $N_0\in{\Bbb N}$ such that
$\bigcap_{N>N_0}^\infty{ dom}(f_N)$ and
$\bigcap_{N>N_0}^\infty{ dom}(\beta_N)$ are nonempty, then the
following limits exist pointwise almost everywhere
\begin{eqnarray}
& &\lim_{N \too \infty} f_N(\beta) \equiv f (\beta)~,~~~{\rm for}~~~
\beta\in\bigcap_{N>N_0}^\infty{ dom}(f_N)
\nonumber \\
& &\lim_{N \too \infty} \beta_N(\vb) \equiv \beta (\vb))~,~~~{\rm for}~~~
\vb\in\bigcap_{N>N_0}^\infty{ dom}(\beta_N)~.
\end{eqnarray}
\end{proposition}

\begin{pro}
See Ref.\cite{ruelle}, chapter 3.4.
\qed
\end{pro}
Henceforth, we shall use $V$ instead of $V_N$ if no explicit reference the
$N$-dependence of $V$ is necessary.

\section{Main Theorem}
\label{mainthm}
In this Section we prove the following theorem:

\begin{theorem}[Necessity condition for Phase Transitions]
Let $V_N$ be a standard, smooth,
confining, short-range potential bounded from below (Definitions \ref{st-pot},
\ref{sr-pot}, \ref{stab-pot} and \ref{conf-pot})
\begin{eqnarray}
        V_N:& & {\cal B}_N\subset{\Bbb R}^N \rightarrow {\Bbb R}\nonumber\\
    V_N(q) &=& \sum_{{\underline i},{\underline j}\in{\cal I}\subset{\Bbb N}^d}
 C_{{\underline i} {\underline j}}
    \Psi (\| \vec{q}_{\underline i} - \vec{q}_{\underline j}  \|) +
    \sum_{{\underline i}\in{\cal I}\subset{\Bbb N}^d}
     \Phi ( \vec{q}_{\underline i} )
\end{eqnarray}
Let $(\Psi ,\Phi)$ be real valued one variable functions, let
${\underline i},{\underline j}$ label interacting pairs of degrees of freedom
within a short-range, and let
$\left \{ \Si_v \right \}_{v\in{\Bbb R}}$ be the family of
$N-1$-dimensional equipotential hypersurfaces
$\Si_v:=V_N^{-1}(v)$, $v\in{\Bbb R}$, of $\Ree^N$.

Let $\vb_0,\vb_1 \in \Ree$, $\vb_0 < \vb_1$. If there exists $N_0$ such that
for any $N>N_0$ and for any $\vb,\vb' \in I_{\vb}=[\vb_0,\vb_1]$
\begin{equation}
\Si_{N\vb}~is~ C^\infty-{\rm diffeomorphic}~~to~ \Si_{N\vb'}, \nonumber
\end{equation}
(notation: $\Si_{N\vb}\approx \Si_{N\vb'}$) then the limit entropy
$S(\vb )$ is of differentiability class
${\cal C}^3(I_{\vb})$, and, consequently, $\beta(\vb )$ belongs to
${\cal C}^2(I_{\vb})$, whence the limit Helmholtz free energy function
$F_\infty \in {\cal C}^2(\stackrel{o}{I}_\beta )$, where
$\stackrel{o}{I}_\beta$  denotes open interior of
$\beta([\vb_0,\vb_1])$), so that the system described by $V$
has neither first nor
second order phase transitions in the inverse-temperature interval
$\stackrel{o}{I}_\beta$.
\label{MainThm}
\end{theorem}
\medskip
The idea of the proof of the Theorem 1 is the following.
In order to prove that a {\it topology change} of the equipotential
hypersurfaces
$\Sigma_v$ of configuration space is a {\it necessary} condition for a
thermodynamic phase transition to occur, we shall prove the {\it equivalent
proposition} that if any two hypersurfaces $\Sigma_{v(N)}$ and
$\Sigma_{v^\prime (N)}$ with $v(N), v^\prime (N) \in (v_0(N),v_1(N))$ are
{\it diffeomorphic} for all $N$, possibly greater than some finite $N_0$,
then {\it no phase transition}
can occur in the (inverse) temperature interval
$[\lim_{N\rightarrow\infty}\beta (\vb_0(N)),\lim_{N\rightarrow\infty}
\beta (\vb_1(N))]$.
To this purpose we have to show that, in the limit $N\rightarrow\infty$ and
$vol (\Lambda^d )/N\ =\ const$, the Helmoltz free energy $F_\infty (\beta ;  H)$
is at least twice differentiable as a function of $\beta =1/T$ in the interval
$[\lim_{N\rightarrow\infty}\beta (\vb_0(N)),\lim_{N\rightarrow\infty}
\beta (\vb_1(N))]$.
For the standard Hamiltonian systems that we consider throughout this paper,
being $F_N(\beta)=-(2\beta)^{-1}\log(\pi/\beta)- f_N(\beta)/\beta$,
this is equivalent to show that the sequence of configurational free energies
$\{ f_N(T; H)\}_{N\in{\Bbb N}_+}$ is {\it uniformly convergent} at
least in
${\cal C}^2$ so that also $\{ f_\infty(T; H)\}\in {\cal C}^2$.

\medskip

We shall give the proof of Theorem 1 through the following Lemmas,
which are separately proven in  subsequent Sections.
\medskip
\begin{lemma}[Absence of critical points]
Let $f: M\rightarrow [a,b]$ a smooth map on a compact manifold $M$
with boundary, such that its Hessian is non-degenerate.
Suppose $f(\partial M)=\{ a, b\}$
and that for any $c,d\in [a,b]$ it is $f^{-1}(c)\approx f^{-1}(d)$, that is
all the level surfaces of $f$ are diffeomorphic.
Then $f$ has no critical points, that is $\Vert\nabla f\Vert\geq C>0$,
in $[a,b]$; $C$ is a constant.
\label{regular-interval}
\end{lemma}

\begin{pro}
\label{proofLm1}

Since $f$ is a good Morse function, let us consider the case of the existence
of -- at least -- one critical value $c\in [a,b]$ so that $\nabla f =0$ at
some points of the level set $f^{-1}(c)$. The set of critical points
$\sigma (c)=\{x_c^{i,k_i}\in f^{-1}(c)\vert (\nabla f)(x_c^{i,k_i})=0\}$ is a
point set \cite{palais}, the index $i$ labels the different critical points and
$k_i$ is the Morse index of the $i$-th critical point. After the ``non-critical
neck'' theorem \cite{palais}, we know that the level sets $f^{-1}(v)$ with
$v\in [a, c-\varepsilon]$ and arbitrary $\varepsilon >0$ are diffeomorphic
because in the absence of critical points in the interval $[a, c-\varepsilon]$
for any $v, v^\prime\in [a, c-\varepsilon]$, with arbitrary $\varepsilon >0$,
$f^{-1}(v)$ is a deformation retraction of $f^{-1}(v^\prime)$ through the
flow associated with the vector field \cite{hirsch}
$X = - \nabla f/\Vert\nabla f\Vert^2$.
Now, in the neighborhood of each critical point $x_c^{i,k_i}$, the existence of
the Morse chart \cite{hirsch} allows to represent the function $f$ as follows
\begin{equation}
f(x)=f(x_c^{i,k_i}) - x_1^2-\dots -x_{k_i}^2+x_{{k_i}+1}^2+\dots +x_n^2~,
\label{morsechart}
\end{equation}
whence the degeneracy of the quadrics, for $v=c$, entailing that the level set
$f^{-1}(c)$ no longer qualifies as a differentiable manifold. Thus for any
$v\in [a, c-\varepsilon]$ and arbitrary $\varepsilon >0$, it is
\begin{equation}
f^{-1}(v) \not\approx f^{-1}(c)~.
\end{equation}
In conclusion, if for any pair of values $v, v^\prime\in [a, b]$
one has $f^{-1}(v^\prime)\approx f^{-1}(v)$, no critical point of $f$ can
exist in the interval $[a, b]$.~\qed

\end{pro}


\begin{lemma}[Smoothness of the structure integral]
Let $V_N$ be a standard, short-range, stable and confining potential function
bounded below. Let
$\left \{ \Si_v \right \}_{v\in{\Bbb R}}$ be the family of
$(N-1)$-dimensional equipotential hypersurfaces
$\Si_v:=V_N^{-1}(v)$, $v\in{\Bbb R}$, of ${\Bbb R}^N$, then we have:
\begin{equation}
If~ for~ any~~v,v'~\in~[v_0,v_1],~\Si_v~\approx~\Si_{v'}
~then ~ \isn~\in ~{\cal C}^\infty(]v_0,v_1[ )\, .
 \nonumber
\end{equation}
\label{isn-cinf}
\end{lemma}
\begin{pro}
The proof of this Lemma is given  in Section \ref{proofLm2}.
\qed
\end{pro}

\begin{lemma}[Uniform convergence]
Let $U$ and $U'$ be two open intervals of $\Ree$. Let $h_N$ be a sequence
of functions from $U$ to $U'$, differentiable on $U$, and
let $h : U \too U'$ be such that
$for~any~ x\in U,\ \lim_{N \rightarrow \infty} h_N(x) = h(x)$.
\newline
If  $there~exists~ M\in \Ree$ such that
$for~ any~ N\in{\Bbb N}~and ~for~any~a \in U~it~is~\\
\left \vert \dyy\frac{dh_N}
{dx}(a) \right \vert \leq M$,
then  $h$ is continuous at $a$ for any $a \in U$.
\label{cor-ascoli}
\end{lemma}
\begin{pro}
From the assumption that  for any $N\in{\Bbb N}$ and for any $a \in U$ it is
$\vert h^\prime_N(a)\vert\leq M$, and after the fundamental theorem of
calculus, the set of functions $\{ h_N\}_{N\in{\Bbb N}}$ is equilipschitzian
and thus uniformly equicontinuous \cite{schwartz}.
Then, from the Ascoli theorem on
equicontinuous sets of applications \cite{schwartz}, it follows
that for any $a \in U$ the closure of the set of functions
$\{ h_N\}_{N\in{\Bbb N}}$ is equicontinuous, and thus the limit function
$h$ is continuous at $a$ for any $a\in U$.~\qed
\end{pro}

\begin{lemma}[Uniform upper bounds]
Let $V_N$ be a standard, short-range, stable and confining potential function
bounded below.
Let
$\left \{ \Si_v \right \}_{v\in{\Bbb R}}$ be the family of
$(N-1)$-dimensional equipotential hypersurfaces
$\Si_v:=V_N^{-1}(v)$, $v\in{\Bbb R}$, of ${\Bbb R}^N$,  if
    \bena
    for~ any~ N,~for~any~\vb,\vb' \in {I_{\vb}}=[\vb_0,\vb_1],~~
      \Si_{N \vb}~\approx~\Si_{N \vb'} \nonumber
    \eena
then
    \bena
    \sup_{N,\vb\in I_{\vb}} \left\vert S_N({\vb})\right\vert
        < \infty~~~{\it and}~~~
    \sup_{N,\vb\in I_{\vb}} \left\vert \frac{\de^k
    S_N}{\de {\vb}^k}({\vb})\right\vert < \infty,~~k=1,2,3,4. \nonumber
    \eena
\label{derivees-majorees}
\end{lemma}
\begin{pro}
The proof of this Lemma is given  in Section \ref{proofLm4}.\qed
\end{pro}

\medskip
\begin{pro} [Theorem 1]
Under the hypothesis that all the level surfaces of $V_N$ are diffeomorphic
in the interval $I_{\vb}$ we know from Lemma \ref{regular-interval} that there
are no critical points of $V_N$ in $I_{\vb}$, i.e. there exists $C(N)>0$ such
that for any $N>N_0$
\begin{eqnarray}
{\rm ~for~} \vb \in I_{\vb},~~{\rm and ~for~any~}
x \in \Sigma_{N\vb},~~\Vert \nabla V_N (x) \Vert \geq C>0 \, .
\label{critpoint}
\end{eqnarray}
Therefore, the restriction of $V_N$
\begin{equation}
{\tilde V}_N =  V_{\vert V_N^{-1}(I_{N\vb})} : V_N^{-1}(I_{N\vb})
\subset B\rightarrow {\Bbb R}
\end{equation}
always defines a Morse function, since $V_N$ is bounded below.
Notice that
\begin{equation}
S_N(\bullet \ ; V_N)_{\vert\stackrel{o}{I}_{\vb}}\equiv
S_N(\bullet \ ; {\tilde V}_N)_{\vert\stackrel{o}{I}_{\vb}}~,
\end{equation}
in what
follows we shall drop the tilde and $V_N$ will denote the above
given restriction.

Now, since the condition (\ref{critpoint}) holds for the hypersurfaces
$\{ \Sigma_{N{\vb}}\}_{{\vb}\in\stackrel{o}{I}_{\vb}}$, from
Lemma \ref{isn-cinf} it follows that for any $N>N_0$,
$\Omega (N{\vb},N)$ is actually in $C^\infty (\stackrel{o}{I}_{\vb})$, where
$\stackrel{o}{I}_{\vb}=(\vb_0, \vb_1)$; this implies that
for any $N>N_0$, also $S_N$
belongs to ${\cal C}^\infty(\stackrel{o}{I}_{\vb})$.

While at any finite $N$ -- under the main  assumption of the theorem --
the entropy functions $S_N$ are smooth, we do not know what happens in the
$N\rightarrow\infty$ limit. To know the behaviour at the limit,
we have to prove the uniform
convergence of the sequence $\{S_N\}_{N\in{\Bbb N}_+}$.
Lemmas \ref{cor-ascoli} and
\ref{derivees-majorees} prove exactly that this sequence is uniformly
convergent  at least in the space
${\cal C}^3(\stackrel{o}{I}_{\vb})$, so that we can conclude that also
$S \in {\cal C}^3(\stackrel{o}{I}_{\vb})$.

As $S=S^{(-)}$ in $I_{\vb}$ (Proposition \ref{pro-elcc}), also
$S^{(-)}$ lies in ${\cal C}^3(\stackrel{o}{I}_{\vb})$ and $\beta$ in
${\cal C}^2(\stackrel{o}{I}_{\vb})$.

Moreover, by definition and existence of the uniform
limit of $\{S_N\}_{N\in{\Bbb N}_+}$, for any
$\vb \in \stackrel{o}{I}_{\vb}$ we can write
\begin{equation}
S(\vb) = f(\beta(\vb)) + \beta(\vb) \cdot \vb \,
\nonumber
\end{equation}
which entails $f \in {\cal C}^2(\beta(\stackrel{o}{I}_{\vb}))$
$\equiv {\cal C}^2(\stackrel{o}{I}_{\beta})$.

Since the kinetic energy term of the Hamiltonian describing the system
${\cal S}$ gives only a smooth contribution, also the Helmoltz  free energy
$F_\infty$ has differentiability class
${\cal C}^2(\stackrel{o}{I}_\beta)$.
Hence we conclude that the system ${\cal S}$ does not undergo neither first
nor second order phase transitions in the inverse-temperature interval
$\beta \in \stackrel{o}{I}_\beta$. \qed
\end{pro}

\begin{corollary}
 Under the same hypotheses of Theorem \ref{MainThm}, let
$\{ M_v\}_{v\in{\Bbb R}}$ be the family of the $N$-dimensional subsets
$M_v:=V_N^{-1}((-\infty, v])$, $v\in{\Bbb R}$, of ${\Bbb R}^N$.
Let $\vb_0,\vb_1 \in \Ree$, $\vb_0 < \vb_1$. If there exists $N_0$ such that
for any $N>N_0$ and for any $\vb,\vb' \in I_{\vb}=[\vb_0,\vb_1]$
\begin{equation}
M_{N\vb}~is~ C^\infty-{\rm diffeomorphic}~~to~ M_{N\vb'}, \nonumber
\end{equation}
then the limit entropy
$S^{(-)}(\vb )$ is of differentiability class
${\cal C}^3(I_{\vb})$, and, consequently,
$\beta(\vb )=\partial S^{(-)}/\partial \vb $ belongs to
${\cal C}^2(I_{\vb})$, whence the limit Helmholtz free energy function
$F_\infty \in {\cal C}^2(\stackrel{o}{I}_\beta )$, where
$\stackrel{o}{I}_\beta$  denotes open interior of
$\beta([\vb_0,\vb_1])$), so that the system described by $V$
has neither first nor
second order phase transitions in the inverse-temperature interval
$\stackrel{o}{I}_\beta$.
\label{corolla}
\end{corollary}

\begin{pro}
If for any $\vb,\vb' \in I_{\vb}=[\vb_0,\vb_1]$ it is $M_{N\vb}\approx M_{N\vb'}$,
then after Bott's ``critical-neck theorem'' \cite{bott}, there are no critical
points of $V_N$ in the interval $[\vb_0,\vb_1]$. As a consequence of the absence
of critical points in $[\vb_0,\vb_1]$, after the ``non-critical neck theorem''
\cite{palais} for any $\vb,\vb' \in I_{\vb}=[\vb_0,\vb_1]$ it is
$\Si_{N\vb}\approx \Si_{N\vb'}$. Now Theorem \ref{MainThm} implies
$S(\vb )\in {\cal C}^3(I_{\vb})$, so that using Proposition \ref{pro-elcc}
we have also $S^{(-)}(\vb )\in {\cal C}^3(I_{\vb})$. Then using equation
(\ref{legendre-tras}) we have $f_\infty(\beta )\in {\cal C}^2(I_{\vb})$ and
thus $F_\infty \in {\cal C}^2(\stackrel{o}{I}_\beta )$, so that neither first nor
second order phase transitions can occur in the inverse temperature interval
$\stackrel{o}{I}_\beta = (\partial S^{(-)}/\partial \vb \vert_{\vb =\vb_0},
\partial S^{(-)}/\partial \vb \vert_{\vb =\vb_1})$.
 \qed
\end{pro}




\section{Proof of Lemma 2, smoothness of the structure integral}
\label{proofLm2}
\pagestyle{headings}

We make use of the following Lemma
\begin{lemma}
Let $U$ be a bounded open subset of $\Ree^N$, let $\psi$ be a Morse function
defined on $U$, $\psi:U \subset \Ree^N \too \Ree$ and ${\cal F}=\{ \Si_v \}_v$
the family of hypersurfaces defined as
$\Sigma_v=\{ x \in U \vert \psi(x)=v \}$, then we have:
\begin{eqnarray}
&&if ~for~any~ v, v' \in [v_0,v_1],~\Si_v~\approx~\Si_v' \nonumber\\
&&then,~for~any~ g \in {\cal C}^\infty (U),~
\int_{\Si_v} g~d\sigma~
{\it is}~{\cal C}^\infty~{\it in}~]v_0,v_1[ \, .\nonumber
\end{eqnarray}
\label{F-cinf}
\end{lemma}

\begin{pro} To prove this Lemma we need the following
Theorem\cite{federer,laurence}:
\smallskip

{\bf Theorem\ (Federer, Laurence).\ } {\it
Let $O \subset \Ree^p$ be a bounded open set. Let $\psi \in {\cal C}^{n+1}
(\bar O)$
be constant on each connected component of
the boundary $\partial O$ and $g \in {\cal C}^n (O)$.

By introducing $O_{t,t'}=\{x \in O\mid t<\psi (x) <t' \}$, and
$F(v)= \int_{ \{ \psi=v \} } g~d\sigma^{p-1}$, where $d\sigma^{p-1}$
represents the Lebesgue measure of dimension $p-1$.

If $C>0$ exists such that $for~any~ x \in O_{t,t'}, \Vert \nabla \psi (x)
\Vert \geq C$, $for~ any~k~s.t.~ 0 \leq k \leq n,~for~any~ v \in ]t,t'[$,
one has
\begin{equation}
\frac{d^k F}{dv^k}(v)
=\int_{\{ \psi=v \}} A^k g~d\sigma^{p-1} \, .
\label{federerderiv}
\end{equation}
with $A g=\nabla \lpt \frac{\nabla \psi}{\| \nabla \psi \|} g \rpt \frac{1}
{\| \nabla \psi \|}$~. }

By applying this Theorem to the function $\psi$ of the
Lemma \ref{F-cinf} we have that, if there exists a constant $C>0$
such that $for~any~ x \in O_{v_0,v_1}$ it is
$\Vert \nabla \psi (x) \Vert \geq C$, then
\begin{equation}
\frac{d^k F}{dv^k}(v)=\int_{\Si_v} A^k g d\sigma,
~~\forall v \in ]v_0,v_1[ \nonumber
\end{equation}

Now, under the hypothesis that
$for~any~ v,v' \in~[v_0,v_1],~\Si_v~\approx~\Si_{v'}$,
we know from Lemma \ref{regular-interval}, ``absence of critical points'',
that this hypothesis is equivalent to the assumption
that $for~any~ v \in~[v_0,v_1], \Si_v$ has no critical points.
Hence there exists a constant $C>0$ such that
$\forall x \in O_{v_0,v_1}$ $\Vert \nabla \psi (x) \Vert \geq C$.
Furthermore, as
$\Vert\nabla \psi \Vert$ is strictly positive, $A$ is a continuous operator
on $O_{v_0,v_1}$.
Thus, being $\Si_v$  compact, $\dyy\frac{d^k F}{dv^k}$ is continuous on the
interval $]v_0,v_1[$, $\forall k$, namely $ \int_{\Si_v} g d\sigma
\in {\cal C}^\infty(]v_0,v_1[)$~.
\medskip

To conclude the proof of the Lemma \ref{isn-cinf} we have to use Lemma
\ref{F-cinf} taking $\psi = V_N$ and $g = 1 / \Vert\nabla V_N\Vert$,
assuming that $V_N$ is a Morse function
and that $\Vert\nabla V_N\Vert$ is strictly positive
(absence of critical points of $V_N$
stemming from the hypothesis of diffeomorphicity of Theorem 1).
\qed
\end{pro}


\section{Proof of Lemma \ref{derivees-majorees}, upper bounds}
\label{proofLm4}
\pagestyle{headings}

The proof of this Lemma is splitted into two parts.
In part A some preliminary results
to be used in part B are given, and in part B
the inequalities of the Lemma \ref{derivees-majorees} are proved.

The proof of Lemma \ref{derivees-majorees} is the core of the proof of
Theorem 1. Thus, as the proof
of Lemma \ref{derivees-majorees} is lengthy, in order to ease its reading
we premise a summary of it.

\noindent{\it Sketch of the proof\ .}

In order to prove Theorem 1, we have to show that the assumption of
diffeomorphicity among the $\Sigma_{N\vb}$ for $\vb\in [\vb_0, \vb_1]$, entails
that $S_\infty (\vb)$ is three times differentiable.
After the Ascoli theorem \cite{schwartz}, this is proved by showing
that for $\vb\in I_{\vb}=[\vb_0, \vb_1]$
and {\it for any} $N$, the function
$S_N(\vb)$ and its first four derivatives are uniformly bounded in $N$ from
above, that is, for any $N\in{\Bbb N}$ and $\vb \in [\vb_0, \vb_1 ]$
\begin{equation}
    \sup \left\vert S_N({\vb})\right\vert
        < \infty \, , \,
 ~\sup\left\vert \frac{\de^k
    S_N}{\de {\vb}^k}
\right\vert < \infty\, ,~k=1,..,4 .
\label{bounds}
\end{equation}

After {\it Definition 1} for the entropy, the first four derivatives of $S_N(\vb)$ are
%
\[
\partial_{\vb}S_N=(1/N)(d v/d \vb)\Omega^{\prime}/\Omega\ ,
\]
\be
\partial^2_{\vb}S_N=N[\Omega^{\prime\prime}/\Omega -
(\Omega^\prime/\Omega)^2]\ , \label{derOm}
\ee
\[
\partial^3_{\vb}S_N=N^2
[\Omega^{\prime\prime\prime}/\Omega -
3\Omega^{\prime\prime}\Omega^\prime/\Omega^2
+2(\Omega^\prime/\Omega)^3]\ ,
\]
\[
\partial^4_{\vb}S_N=N^3[ \Omega^{iv}/\Omega -4\Omega^{\prime\prime\prime}
\Omega^\prime/\Omega^2 -3(\Omega^{\prime\prime}/\Omega )^2 +12
\Omega^{\prime\prime}(\Omega^\prime)^2/\Omega^3
-6(\Omega^\prime/\Omega)^4]\ ,
\]
where the prime indexes stand
for derivations of $\Omega(v,N)$ with respect to $v=\vb N$.
In order to verify whether the conditions  (\ref{bounds}) are fulfilled, we
must be able to estimate the $N$-dependence of all the addenda in these
expressions for the derivatives of $S_N$.

Being the assumption of diffeomorphicity of the  $\Sigma_{N\vb}$ equivalent
to the absence of critical points of the potential, we can
use the derivation formula \cite{federer,laurence}
    \bena
\frac{d^k}{dv^k}\isn = \int_{\Si_v}\ngV \ A^k \lpt \unv \rpt \frac{d \sigma}
                       {\ngV} \dd ,
    \label{derfed}
    \eena
where $A^k$ stands for $k$ iterations of the operator
\[
A(\bullet ) =\nabla \lpt \frac{\nabla V}{\| \nabla V \|}\ \bullet \rpt \frac{1}
{\| \nabla V \|}~.
\]
A technically crucial step to prove the Theorem is to use the above formula
(\ref{derfed}) to compute the derivatives of $\Omega (v,N)$, in fact these are
transformed into the surface integrals
of explicitly computable combinations and powers of a few basic ingredients,
like
$\Vert\nabla V\Vert$, $\partial V/\partial q_i$,
$\partial^2 V/\partial q_i\partial q_j$,
$\partial^3 V/\partial q_i\partial q_j\partial q_k$ and so on.

The first uniform bound in Eq.(\ref{bounds}), $\vert S_N({\vb})\vert < \infty$,
is a simple consequence of the intensivity of $S_N({\vb})$.

To prove the boundedness of the first derivative of $S_N$, we compute
its expression by means of the first of Eqs.(\ref{derOm}) and of Eq.(\ref{derfed}),
which reads
\begin{equation}
\frac{\de S_N}{\de {\vb}}=\frac{1}{\Omega}\int_{\Sigma_{\vb N}}\left[
\frac{\Delta V}{\ngV^2}
  -2 \frac{\sum_{i,j}
\partial^i V\partial^2_{ij} V \partial^j V}{\ngV^4}\right]
\frac{d\sigma}{\ngV}~,
\label{der1}
\end{equation}
with $\partial_iV=\partial V/\partial q^i$ and $i,j=1,\dots,N$, whence
(with an obvious meaning of $\langle\cdot\rangle_{\Sigma_v}$)
\begin{equation}
\left\vert\frac{\de S_N}{\de {\vb}}\right\vert \leq
\left \langle \frac{\mid \Delta V \mid}{ \ngV^2}\right \rangle_{\Sigma_v}
  +2\left \langle \frac{\left\vert \sum_{i,j}
\partial^i V\partial^2_{ij} V \partial^j V
\right \vert }{\ngV^4}\right \rangle_{\Sigma_v},
\label{ineq}
\end{equation}
the r.h.s. of this inequality -- in the absence of critical points of the
potential -- can be bounded from above by (see Lemma \ref{comportement_ngV})
\be
\frac{\left \langle \mid \Delta V \mid \right \rangle_{\Sigma_v}}
{\langle \ngV^2 \rangle_{\Sigma_v}}
\ + O\left(\frac{1}{N}\right)
+2 \frac{\left \langle \sum_{i,j=1}^N \mid \partial^i V\partial^2_{ij} V \partial^j V \mid
\right \rangle_{\Sigma_v}}{ \langle \ngV^4\rangle_{\Sigma_v}}\
+ O\left(\frac{1}{N^2}\right)\ . \label{ineq1}
\ee
As we have assumed that $V$ is smooth and bounded below, and after the argument
put forward in Remark \ref{fattorizza}, we have
 $\langle \mid \Delta V \mid \rangle_{\Sigma_v}
=\langle \mid \sum_{i=1}^N\partial^2_{ii} V \mid \rangle_{\Sigma_v}
\leq N\max_i \langle  \mid \partial_{ii}^2 V \mid
 \rangle_{\Sigma_v}$ and, as we have also assumed that $V$ is a short range
potential, the number of non-vanishing matrix elements $\partial^2_{ij}V$ is
$N(n_p+1)$ where  $n_p$ is the number of neighbouring particles in the
interaction range of the potential, thus
$\left \langle \mid \partial^i V\partial^2_{ij} V \partial^j V
\mid\right\rangle_{\Sigma_v} \leq  N (n_p+1)\max_{i,j} \langle \mid \partial^i V
\partial^2_{ij} V \partial^j V \mid\rangle_{\Sigma_v}$.

Moreover, the following lower bounds exist for the denominators in the inequality
(\ref{ineq1}):

\noindent$\langle \ngV^2 \rangle_{\Sigma_v}
\geq N~\min_{i} \langle \left( \partial_i V \right)^2 \rangle_{\Sigma_v}$, and
$\langle \ngV^4 \rangle_{\Sigma_v}
\geq N^2~\min_{i,j}
\langle \left( \partial_i V \right)^2 \left( \partial_j V \right)^2
 \rangle_{\Sigma_v}$.

Finally, putting $ m=\max_{i,j} \langle \mid \partial^i
 V \partial^2_{ij} V \partial^j V \mid\rangle_{\Sigma_v}$,
$c_1=\min_{i} \langle \left( \partial_i V \right)^2
\rangle_{\Sigma_v}$ and $c_2=\min_{i,j}
\langle \left( \partial_i V \right)^2 \left( \partial_j V \right)^2
\rangle_{\Sigma_v}$, by substituting in Eq.(\ref{ineq1}) the upper bounds
for the numerators and the lower bounds for the denominators we obtain
\be
\left\vert\frac{\de S_N}{\de {\vb}}\right\vert \leq
\frac{\max_i \langle  \mid \partial_{ii}^2 V \mid
 \rangle_{\Sigma_v}}{c_1}
+ O\left( \frac{1}{N} \right)\
+2 \frac{n_p~m}{c_2N}+ O \left( \frac{1}{N^2} \right)\
\label{maggior}
\ee
which, in the limit $N\rightarrow\infty$, shows
that the first derivative of the entropy is uniformly
bounded by a finite constant. This first step proves that $S_\infty (\vb )$
is continuous.

The three further steps, concerning boundedness of the higher order
derivatives, involve similar arguments to be applied to a number of terms
which is rapidly increasing with the order of the derivative.
But many of these terms can be grouped in the form of the variance or
higher moments of certain quantities, thus allowing the use of a powerful
technical trick to compute their $N$-dependence.
For example, using Eq.(\ref{derfed}) in
the expression for $\partial^2_{\vb}S_N$, we get
 \be
  \left | \frac{\de^2 S_{N}}{\de {\vb}^2} \right |
  \leq N \Big |
  \langle\alpha ^2\rangle_{\Sigma_v}\!
  - \langle\alpha\rangle_{\Sigma_v}^2\Big |
  + N \Big |  \langle
  \psvd \ps \lpt \alpha \rpt
  \rangle_{\Sigma_v} \Big |
  \label{dersec}
  \ee
where $\alpha ={\ngV\ A (1/\ngV )}$ and $\ps =\nabla/\ngV$. Now,
it is possible to think of the scalar function $\alpha$ as if it were
a random variable, so that
the first term in the r.h.s. of Eq.(\ref{dersec}) would be its second moment.
Such a possibility is related with the general validity of the Monte Carlo
method to compute multiple integrals.
In particular, since the $\Sigma_v$ are smooth, closed ($V$ is non-singular),
without critical points and representable as the union of suitable subsets
of ${\Bbb R}^{N-1}$, the standard Monte Carlo method  \cite{mcmc}
is applicable to
the computation of the averages $\langle\cdot\rangle_{\Sigma_v}$ which become
sums of standard integrals in ${\Bbb R}^{N-1}$. This means
that a random walk can be
constructively defined on any $\Sigma_v$, which conveniently samples the
desired measure on the surface (see Lemma 6).
Along such a random walk, usually called Monte Carlo Markov
Chain (MCMC), $\alpha$ and its powers
behave as random variables whose ``time'' averages along the MCMC converge to
the surface averages $\langle\cdot\rangle_{\Sigma_v}$.
Notice that the actual computation of these surface averages goes beyond
our aim, in fact, we do not need the numerical values -- but only the
$N$-dependences -- of the upper bounds of the derivatives of the entropy.
Therefore, all what we need is just knowing that in principle a suitable MCMC
exists on each $\Sigma_v$.
Now, the function $\alpha$ is the integrand in square brackets in
Eq.(\ref{der1}), where the second term vanishes at large $N$, as is clear from
Eq.(\ref{maggior}). Therefore, at increasingly large $N$, the approximate
 expression
$\alpha =\sum_{i=1}^N \partial^2_{ii}V/\Vert\nabla V\Vert^2$ tends to become
exact. $\alpha$ is in the form of a sum function
$\alpha  =N^{-1} \sum_{i =1}^N a_i$ of terms
$a_i=N \partial^2_{ii}V/\Vert\nabla V\Vert^2$, of ${ O}(1)$ in $N$, which,
along a MCMC, behave as
independent random variables with probability densities $u_i (a_i)$ which
we do not need to know explicitly.
Then, after a classical ergodic theorem for sum functions, due to Khinchin
\cite{khinchin},
based on the Central Limit Theorem of probability theory, $\alpha$ is a
gaussian-distributed random variable; as its variance decreases linearly
with $N$,
 $\lim_{N\to\infty}  N  | \langle {\alpha}^2\rangle_{\Sigma_v}\!
-  \langle {\alpha} \rangle_{\Sigma_v}^2 | = const <\infty$.

Arguments similar to those above used for the first derivative of $S_N$
lead to the result
$\lim_{N\to\infty}  N  |  \langle\psvd \ps \lpt \alpha \rpt\rangle_{\Sigma_v}
|= const <\infty$, which, together with what has been just found for
the variance
of $\alpha$, proves the uniform boundedness also of the second derivative of
$S_N$ under the hypothesis of diffeomorphicity of the $\Sigma_v$.

Similarly, but with an increasingly tedious work, we can treat the
third and fourth derivatives of the entropy. In fact, despite the large number
of terms contained in their expressions, they again belong only to two
different categories: those terms which can be grouped in the form of higher
moments of the function $\alpha$, and whose $N$-dependence is known after
the above mentioned theorem due to Khinchin and Lemma 7, and those terms whose
$N$-dependence can be found by means of the same kind of estimates
given above for $\partial_{\vb}S_N$. Eventually, after a lenghty but rather
mechanical work, also the third and fourth derivatives of $S_N$ are shown
to be uniformly bounded as prescribed by Eq.(\ref{bounds}).
Whence the proof of Theorem 1.

\medskip

\subsection{Part A}

We begin by showing that on any $(N-1)$-dimensional
hypersurface $\Sigma_{N\vb}=V^{-1}_N(N\vb )
=\{X\in{\Bbb R}^{N}\ \vert \ V_N(X) =N\vb\}$ of ${\Bbb R}^{N}$, we can define
a homogeneous non-periodic random Markov chain whose probability measure is
the configurational microcanonical measure, namely
$d\sigma /\Vert\nabla V_N\Vert$.

Notice that at any finite $N$ and in the absence of critical points of the
potential $V_N$ (because of $\Vert\nabla V_N\Vert \geq C >0$) the
microcanonical measure is smooth. The microcanonical averages
$\langle~\rangle_{N,v}^{\mu c}$ are then equivalently computed as  ``time''
averages along the previously mentioned Markov chains.

In the following, when no ambiguity is possible, for the sake of notation we
shall drop the suffix $N$ of $V_N$.

\begin{lemma}
\label{mesure_ergodique}
On each finite dimensional level set $\Sigma_{N\vb}=V^{-1}(N\vb )$ of a
standard, smooth, confining, short range potential $V$ bounded below, and in
the absence of critical points, there
exists a random Markov chain of points
$\{X_i\in{\Bbb R}^{N}\}_{i\in{\Bbb N_+}}$, constrained by the condition
$V(X_i) = N{\vb}$, which has
\begin{equation}
d\mu =\frac{d\sigma}{\Vert\nabla V\Vert} \left(\int_{\Sigma_{N\vb}}
\frac{d\sigma}{\Vert\nabla V\Vert}\right)^{-1}
\label{prob}
\end{equation}
as its probability measure, so that, for a smooth function
$F :{\Bbb R}^{N}\rightarrow{\Bbb R}$ it is
\begin{equation}
 \left(\int_{\Sigma_{N\vb}}
\frac{d\sigma}{\Vert\nabla V\Vert}\right)^{-1}
\int_{\Sigma_{N\vb}}\frac{d\sigma}{\Vert\nabla V\Vert}\ F =
\lim_{n\rightarrow\infty}\frac{1}{n}\sum_{i=1}^n F(X_i)~.
\label{integrale}
\end{equation}
\end{lemma}

\begin{pro}

As the level sets $\{\Sigma_{N\vb}\}_{\vb\in{\Bbb R}}$ are compact
codimension-$1$ hypersurfaces of ${\Bbb R}^{N}$, there exists on each of them
a partition of unity \cite{thorpe}.
Thus, denoting by $\{U_i\}$, $1\leq i \leq m$, an arbitrary finite
covering of $\Sigma_{N\vb}$ by means of domains of coordinates (for example
by  means of open balls), a set of smooth functions  $\{\varphi_i\}$
exists, with $1\geq\varphi_i\geq 0$
and $\sum_i\varphi_i =1$, for any point of $\Sigma_{N\vb}$.
Since the hypersurfaces $\Sigma_{N\vb}$ are compact and oriented,
the partition of the unity $\{\varphi_i\}$ on $\Sigma_{N\vb}$, subordinate
to a collection $\{U_i\}$ of one-to-one local parametrizations of
$\Sigma_{N\vb}$, allows to represent the integral of a given smooth
$(N-1)$-form $\omega$ as follows
\[
\int_{\Sigma_{N\vb}} \omega^{(N-1)} =
\int_{\Sigma_{N\vb}}\left(\sum_{i =1}^m\varphi_i (x)\right)
\omega^{(N-1)}(x)=
\sum_{i =1}^m \int_{U_i} \varphi_i\omega^{(N-1)}(x)~.
\]
Now we proceed constructively by showing how a Monte Carlo Markov Chain (MCMC),
having (\ref{prob}) as its probability measure, is constructed on
a given $\Sigma_{N\vb}$.

We consider sequences of random values $\{x_i : i\in\Lambda\}$, with
$\Lambda$ the finite set of indexes of the elements of the partition of the
unity on $\Sigma_{N\vb}$, and $x_i =(x^1_i, \dots,x^{N-1}_i)$ the
local coordinates with respect to $U_i$ of an arbitrary representative
point of the set $U_i$ itself. Then we define the weight $\pi (i)$ of the
$i$-th element of the partition as
\begin{equation}
\pi (i)=\left( \sum_{k=1}^m \int_{U_k}\varphi_k\
\frac{d\sigma}{\Vert\nabla V\Vert} \right)^{-1}
\int_{U_i}\varphi_i\ \frac{d\sigma}{\Vert\nabla V\Vert}
\label{peso}
\end{equation}
and the transition matrix elements \cite{mcmc}
\begin{equation}
p_{ij} = \min \left[ 1, \frac{\pi (j)}{\pi (i)}\right]
\label{pij}
\end{equation}
which satisfy the detailed balance equation $\pi (i) p_{ij}= \pi (j) p_{ji}$.
Starting from an arbitrary element of the partition, labeled by $i_0$, and
using the transition probability (\ref{pij}) we obtain a random Markov chain
$\{i_0, i_1\dots, i_k, \dots\}$ of indexes and, consequently,  a random
Markov chain of points $\{x_{i_0},x_{i_1},\dots, x_{i_k}, \dots\}$ on the
hypersurface $\Sigma_{N\vb}$.
Now, let $(x^1_P, \dots,x^{N-1}_P)$ be the local coordinates of a point $P$ on
$\Sigma_{N\vb}$ and define a local reference frame as
$\{\partial/\partial x^1_P,\dots,\partial/\partial x^{N-1}_P, n(P)\}$
where $n(P)$ is the
outward unit normal
vector at $P$; through the point-dependent matrix which operates the change
from this basis
to the canonical basis $\{e_1,\dots,e_{N}\}$ of ${\Bbb R}^{N}$ we can
associate to the Markov chain $\{x_{i_0},x_{i_1},\dots, x_{i_k}, \dots\}$
an equivalent chain $\{X_{i_0},X_{i_1},\dots, X_{i_k}, \dots\}$ of points
identified through their coordinates in ${\Bbb R}^{N}$ but still constrained
to belong to the subset $V(X) = v$, that is to $\Sigma_{N\vb}$.
By construction, this Monte Carlo Markov Chain has the probability density
(\ref{prob}) as its invariant probability measure \cite{mcmc}, moreover,
for smooth functions $F$, smooth potentials $V$ and in the absence of critical
points, $F/\Vert\nabla V\Vert$ has a limited variation on each set $U_i$,
thus the
partition of the unity can be made as fine grained as needed -- keeping it
finite -- to make Lebesgue integration convergent, hence
Equation (\ref{integrale}) follows.
\qed
\end{pro}
\medskip

In part B we shall need the $N$-dependence of the momenta, up to the fourth
order, of the sum of a large number $N$ of mutually independent random
variables.
These $N$-dependences are worked out in what follows by using and extending
some results due to Khinchin \cite{khinchin}.

\begin{definition}

Let us consider a sequence $\{ \eta_k \}_{k=1,..,N}$ of mutually independent
random quantities with probability densities $\{ u_k(x) \}_{k=1,..,N}$.
Let us denote with $a_k=\int x\ u_k(x)\ dx$ the mean of the $k$-th quantity
and with
\begin{eqnarray}
b_k=\int (x-a_k)^2~u_k(x)~dx&~~~~&
c_k=\int \vert x-a_k\vert^3~u_k(x)~dx \nonumber \\
d_k=\int (x-a_k)^4~u_k(x)~dx&~~~~&
e_k=\int \vert x-a_k\vert^5~u_k(x)~dx \nonumber
\end{eqnarray}
its higher moments.
\label{prekhinchin}
\end{definition}

{\bf Theorem\ (Khinchin).\ } {\it
Let us consider a sequence $\{ \eta_k \}_{k=1,..,N}$ of mutually independent
random quantities with probability densities $\{ u_k(x) \}_{k=1,..,N}$.
Without any significant loss of generality we assume that the $a_k$ are zero.
Under the conditions of validity of the Central Limit Theorem (see
\cite{khinchin}), the probability density
$U_N(x)$ of $s_N=\sum_{k=1}^N \eta_k$ is given by
\begin{eqnarray}
U_N(x) &=&\frac{1}{(2 \pi B_N)^\frac{1}{2}}
\exp \left [ - \frac{x^2}{2 B_N} \right ] + \frac{S_N + T_N x}
{B_N^\frac{5}{2}} \nonumber\\
&+& O \left ( \frac{1+ \mid x \mid^3}{N^2} \right ),
~~~~~\forall \mid x \mid < 2 \log^2 N \label{unx_1} \\
&&\\
U_N(x) &=& \frac{1}{(2 \pi B_N)^\frac{1}{2}}
\exp \left [ - \frac{x^2}{2 B_N} \right ] + O \left ( \frac{1}{N} \right ),
~~~~~\forall x\in {\Bbb R} \label{unx_2}
\end{eqnarray}
where $B_N= \sum_{i=1}^{N} b_i$ and where $S_N$ and $T_N$ are independent of
$x$ such that  $\lim_{N \too \infty} N^{-1}~S_N$ and
$\lim_{N \too \infty} N^{-1}~T_N$ are finite values (allowed to vanish)
and where $\log^2 N$ stands for $(\log N)^2$.  }
\medskip
\begin{lemma}
Consider a sequence $\{ \eta_k \}_{k=1,..,N}$ of zero mean, mutually
independent, random variables with probability densities
$\{ u_k(x) \}_{k=1,..,N}$. Denote with $B_N'$, $C_N'$ and $D_N'$
the second, third  and fourth moments respectively of
$s_N' = \frac{1}{N} \sum_{k=1}^{N} \eta_k$, and with $K_N'= D_N'-3 {B_N'}^2$
the fourth cumulant of $s_N'$.

If the random quantities fulfil the hypotheses of the Central Limit Theorem,
then
\begin{eqnarray}
(i)~~~~~ \lim_{N \too \infty} N~B_N' &=&cst < \infty \nonumber \\
(ii)~~~~~ \lim_{N \too \infty} N^2~C_N' &=&0 \nonumber \\
(iii)~~~~~ \lim_{N \too \infty} N^3~K_N' &=&0 \nonumber
\end{eqnarray}
\label{m234_borne}
\end{lemma}

\begin{pro}{\it Assertion $(i)$.}

Let $\tilde{B}_N$ be the second moment of $s_N=\sum_{k=1}^{N} \eta_k$. After
the above reported Khinchin theorem, we have
\begin{eqnarray}
\tilde{B}_N &=& \int \mid x \mid^2 \tilde{U}_N(x) dx \nonumber \\
&=&\frac{1}{(2 \pi B_N)^\frac{1}{2}} \int \mid x \mid^2
\exp \left [ - \frac{x^2}{2 B_N} \right ] dx
+ \int \mid x \mid^2 R_N(x) dx \nonumber
\end{eqnarray}
where $R_N(x)$ is a remainder of order $1/N$. The r.h.s. of this equation is
the second moment of the gaussian distribution which is just $B_N$.
Then $\tilde{B}_N$ can be rewritten, using again Khinchin theorem, as
\begin{eqnarray}
\lim_{N \too \infty} \tilde{B}_N&=& \lim_{N \too \infty} B_N +
\lim_{N \too \infty} \int_{\mid x \mid < 2 \log^2 N}
\mid x \mid^2 \frac{S_N + T_N x}{B_N^\frac{5}{2}} \nonumber \\
&=& \lim_{N \too \infty} B_N +
\lim_{N \too \infty} \int_{\mid x \mid < 2 \log^2 N}
\mid x \mid^2 \frac{S_N}{B_N^\frac{5}{2}} \nonumber \\
&=& \lim_{N \too \infty} B_N +
 \frac{2^4}{3} \lim_{N \too \infty}\frac{S_N~\log^6 N}{B_N^\frac{5}{2}}
\nonumber
\end{eqnarray}
Now let $U'_N(x)$ be the probability density of
$s_N' = \frac{1}{N} \sum_{k=1}^{N} \eta_k$, its second moment $B_N'$
is equal to
\begin{eqnarray}
B_N' = \int \mid x \mid^2 U_N'(x) dx = \frac{1}{N^2}~\tilde{B}_N \nonumber
\end{eqnarray}
and thus
\begin{eqnarray}
\lim_{N \too \infty} N~B_N' = \lim_{N \too \infty} \frac{B_N}{N} +
 \frac{2^4}{3} \lim_{N \too \infty}\frac{S_N~\log^6 N}{N~B_N^\frac{5}{2}}~.
\end{eqnarray}
Since $\lim_{N \too \infty} N^{-1}~B_N$ is a finite non-vanishing value
and $\lim_{N \too \infty} N^{-1}~S_N$ is a finite value, we conclude that
\begin{eqnarray}
\lim_{N \too \infty} N~B_N' = cst < \infty~.
\end{eqnarray}
\qed
\end{pro}

\begin{pro}{\it Assertion $(ii)$.}

Let $\tilde{C}_N$ be the third moment of $s_N=\sum_{k=1}^{N} \eta_k$. After
Khinchin theorem we have
\begin{eqnarray}
\tilde{C}_N &=& \int \mid x \mid^3 \tilde{U}_N(x) dx \nonumber \\
&=&\frac{1}{(2 \pi B_N)^\frac{1}{2}} \int \mid x \mid^3
\exp \left [ - \frac{x^2}{2 B_N} \right ] dx
+ \int \mid x \mid^3 R_N(x) dx
\nonumber
\end{eqnarray}
where $R_N(x)$ is a remainder of order $1/N$.
The first term of the r.h.s. is identically vanishing because it is an
odd moment of a gaussian distribution. Thus $\tilde{C}_N$ can be rewritten,
using again Khinchin theorem, as
\begin{eqnarray}
\lim_{N \too \infty} \tilde{C}_N&=& \lim_{N \too \infty}
\int_{\mid x \mid < 2 \log^2 N}
\mid x \mid^3 \frac{S_N + T_N x}{B_N^\frac{5}{2}} \nonumber \\
&=& \lim_{N \too \infty} \int_{\mid x \mid < 2 \log^2 N}
\mid x \mid^3 \frac{S_N}{B_N^\frac{5}{2}}
= 2^3 \lim_{N \too \infty} \frac{S_N~\log^8 N}{B_N^\frac{5}{2}} \nonumber
\end{eqnarray}
Now let $U'_N(x)$ be the probability density of
$s_N' = \frac{1}{N} \sum_{k=1}^{N} \eta_k$, its third moment $C_N'$
is equal to
\begin{eqnarray}
C_N' = \int \mid x \mid^3 U_N'(x) dx = \frac{1}{N^3}~\tilde{C}_N \nonumber
\end{eqnarray}
which leads to the conclusion
\begin{eqnarray}
\lim_{N \too \infty} N^2~C_N'
=  2^3 \lim_{N \too \infty} \frac{S_N~\log^8 N}{N~B_N^\frac{5}{2}}=0~.
\end{eqnarray}
\qed
\end{pro}

\begin{pro}{\it Assertion $(iii)$.}

Let $\tilde{K}_N$ be the fourth cumulant of $s_N=\sum_{k=1}^{N} \eta_k$.
we have
\begin{eqnarray}
\tilde{K}_N=\frac{1}{3} \int x^4 \tilde{U}_N(x) dx
- \left ( \int x^2 \tilde{U}_N(x) dx \right )^2
\end{eqnarray}
which, using  Khinchin theorem, can be written as
\begin{eqnarray}
\tilde{K}_N&=&\frac{1}{3} \int x^4 G_N(x) dx
- \left ( \int x^2 G_N(x) dx \right )^2 \nonumber \\
&+& \frac{1}{3} \int x^4 R_N(x) dx
- \left ( \int x^2 R_N(x) dx \right )^2
-  2~\int x^2 R_N(x) dx ~\int x^2 G_N(x) dx \nonumber
\end{eqnarray}
where $G_N(x)=(2 \pi B_N)^{-\frac{1}{2}} \exp \left [ - \frac{x^2}{2 B_N}
\right ]$ is a gaussian probability distribution and $R_N(x)$ the remainder of
order $1/N$.

The sum of the first two terms of the r.h.s. of the equation above is the
fourth cumulant of a gaussian distribution, thus vanishing.

Again using Khinchin theorem we can write
\begin{eqnarray}
\lim_{N \too \infty} \tilde{K}_N
&=& \frac{1}{3} \lim_{N \too \infty} \int_{\mid x \mid < 2 \log^2 N}
x^4 \frac{S_N + T_N x}{B_N^\frac{5}{2}} dx \nonumber \\
&~~~~-&\lim_{N \too \infty} \left ( \int_{\mid x \mid < 2 \log^2 N}
x^2 \frac{S_N + T_N x}{B_N^\frac{5}{2}} dx \right )^2 \nonumber \\
&~~~~-&\lim_{N \too \infty} \int_{\mid x \mid < 2 \log^2 N}
x^2 \frac{S_N + T_N x}{B_N^\frac{5}{2}} dx~\int x^2 G_N(x) dx \nonumber \\
&=& \frac{2^6}{15} \lim_{N \too \infty} \frac{\log^{10} N~S_N}{B_N^\frac{5}{2}}
- \frac{2^8}{9} \lim_{N \too \infty} \frac{\log^{12} N~S_N^2}{B_N^5}
\nonumber \\
&~~~~-& \frac{2^4}{3} \lim_{N \too \infty} \frac{\log^6 N~S_N}{B_N^\frac{5}{2}}
~.
\end{eqnarray}

Knowing that $\lim_{N \too \infty} N^{-1}~B_N$ is a finite non vanishing value,
that $\lim_{N \too \infty} N^{-1}~S_N$ is a finite value, that
$\int x^2 G_N(x) dx \equiv B_N$, and that
\begin{eqnarray}
K_N' = \frac{1}{3} \int \mid x \mid^4 U_N'(x) dx
- \left ( \int \mid x \mid^2 U_N'(x) dx \right )^2
= \frac{1}{N^4}~\tilde{K}_N \nonumber
\end{eqnarray}
we conclude
\begin{eqnarray}
\lim_{N \too \infty} N^3~K_N' &=&
\frac{2^6}{15} \lim_{N \too \infty} \frac{\log^{10} N~S_N}{N~B_N^\frac{5}{2}}
- \frac{2^8}{9} \lim_{N \too \infty} \frac{\log^{12} N~S_N^2}N~{B_N^5}
\nonumber \\
&~~~~-& \frac{2^4}{3} \lim_{N \too \infty}
\frac{\log^6 N~S_N}{N~B_N^\frac{3}{2}} =0~.\nonumber
\end{eqnarray}

This completes the proof of our Lemma \ref{m234_borne}.
\qed
\end{pro}
\medskip


\begin{remark}
If $V_N$ is a standard, confining, short-range and stable potential, at
large $N$ the entropy function
$S_N(\vb) = \frac{1}{N} \log \Omega \lpt N\vb, N \rpt$
is an intensive quantity, that is
\be
S_{2N} ( \vb) \simeq S_N ( \vb)~.  \nonumber
\ee
This is the obvious consequence of the well known fact that
\be
N S_N(\Lambda^d,\vb) = N_1 S_{N_1}(\Lambda_1^d,\vb)+
N_2 S_{N_2}(\Lambda_2^d,\vb) +{ O} \lpt {\log N} \rpt
\label{separation1}
\ee
which is proved in textbooks\cite{ruelle} and which has also the important
consequence summarized in the following remark.
\label{fn-intensive}
\end{remark}

\begin{remark}
A consequence of equation (\ref{separation1}) is that
\begin{equation}
\Omega (N\vb,N_1+N_2,\Lambda_1^d\cup\Lambda_2^d)=
\Omega (N_1\vb,N_1,\Lambda_1^d)\ \Omega (N_2\vb,N_2,\Lambda_2^d)
\ \theta (N)~,
\end{equation}
where $\theta (N)$ is such that $[\theta (N)]^{1/N}={ O}(N^{1/N})\rightarrow 1$
for $N\rightarrow\infty$.
For two identical subsystems the potential energy
is equally shared among them, with vanishing relative fluctuations in the
$N\rightarrow\infty$ limit.
\label{remarque_factorisation}
\end{remark}

\begin{remark}
In the hypotheses of Theorem 1, $V$ contains only short range interactions and
its functional form does not change with $N$, i.e. the functions $\Psi$ and $\Phi$
in Definitions 3 and 4 do not depend on $N$. In other words, we are tackling physically
homogeneous systems, which, at any $N$, can be considered as the union of smaller and
identical subsystems. At large $N$, if a system is partitioned in a number $k$ of
sufficiently large subsystems, then the generalization to $k$ components of the factorization
of configuration space given in
Remark \ref{remarque_factorisation} holds. Therefore, the averages of functions of interacting
variables, belonging to a given block, do not depend neither on the subsystems where they are
computed (the potential functions are the same on each block after suitable relabeling of the variables),
nor on the total number $N$ of degrees of freedom.
\label{fattorizza}
\end{remark}

\begin{lemma}
Let $\{x_i\}_{i=1,\dots,N}$ and $\{y_i\}_{i=1,\dots,N}$ be two independent sets of
mutually independent non negative random quantities. Define $X=\sum_{i=1}^N x_i$ and
$Y=\sum_{i=1}^N y_i$. Let $Y>0$ for any realisation of the random variables
$\{y_i\}_{i=1,\dots,N}$. Let $\langle X\rangle$, $\langle Y\rangle$ denote the averages
over an arbitrarily
large number of realisations of the sets of random variables
$\{x_i\}_{i=1,\dots,N}$ and $\{y_i\}_{i=1,\dots,N}$, respectively.

In the limit $N\to\infty$, it is
\[
\left\langle \frac{X}{Y}\right\rangle =\frac{\langle X\rangle}{\langle Y\rangle}\ .
\]
\label{comportement_ngV}
\end{lemma}

\begin{pro} After the Khinchin Theorem recalled below Definition \ref{prekhinchin},
in the large $N$ limit both $X$ and $Y$ are gaussian distributed random variables.
Setting $\delta X=X-\langle X\rangle$ and $\delta (1/Y)=1/Y-\langle 1/Y\rangle$ we have
\begin{equation}
\left\langle\frac{X}{Y}\right\rangle =\langle X\rangle\left\langle\frac{1}{Y}
\right\rangle + \left\langle\delta X\ \delta\left(\frac{1}{Y}\right)\right\rangle\ .
\label{newlm8}
\end{equation}
Moreover
\[
\left\langle\delta X\ \delta\left(\frac{1}{Y}\right)\right\rangle\leq
\left\langle\delta Z\ \delta\left(\frac{1}{Z}\right)\right\rangle
\]
where $Z=X$ if $\langle(\delta X)^2\rangle\geq \langle[\delta (1/Y)]^2\rangle$
or $Z=Y$ if $\langle(\delta Y)^2\rangle\geq \langle(\delta X)^2\rangle$, and
\be
\left\langle\delta Z\ \delta\left(\frac{1}{Z}\right)\right\rangle = 1 - 2\langle Z\rangle
\left\langle \frac{1}{Z}\right\rangle + \langle Z\rangle^2\left\langle\frac{1}
{Z^2}\right\rangle\ .
\label{newlm8-2}
\ee
Now, for a gaussian random variable $Z$ such that $\langle Z\rangle >0$, we have
\[
\left\langle\frac{1}{Z}\right\rangle
 =\frac{1}{\langle Z\rangle}\left\langle \frac{1}{1 +(Z-\langle Z\rangle)/
\langle Z\rangle}\right\rangle =\frac{1}{\langle Z\rangle}\left[ 1+
\frac{\langle (Z-\langle Z\rangle)^2\rangle}{3\langle Z\rangle^2} - \cdots\right]
\]
where all the terms with odd powers in the series expansion of
$1/(1+\delta Z/\langle Z\rangle)$
vanish, and the even powers terms are powers of the quadratic term which is
$O(1/N)$, thus in the limit $N\to\infty$
\be
\left\langle\frac{1}{Z}\right\rangle = \frac{1}{\langle Z\rangle}\ .
\label{newlm8-3}
\ee
Using Eq.(\ref{newlm8-3}) in  Eq.(\ref{newlm8-2}) we get
\[
\left\langle\delta X\ \delta\left(\frac{1}{Y}\right)\right\rangle \leq
-1 + \frac{\langle Z\rangle^2}{\langle Z^2\rangle}=O(1/N)\ ,
\]
which, used in Eq.(\ref{newlm8}) together with Eq.(\ref{newlm8-3}), leads to the
final result.
~\qed
\end{pro}

\subsection{Part B}
\label{parteB}
This part is devoted to the proof of the existence of uniform upper bounds
as affirmed in the Lemma \ref{derivees-majorees}.

We shall prove that the {\it supremum} on $N$ and on $\vb \in {I_{\vb}}$ exists
of up to the fourth derivative of $S_N({\vb})$.
The proof of the existence of $sup_N$ will be given by showing that the
functions considered have a finite value in the $N\rightarrow\infty$
limit for any $\vb \in {I_{\vb}}$.
The existence of the {\it supremum} on $\vb$ is then a consequence of
compactness \cite{nota1} of the set ${I_{\vb}}$.
\medskip

\begin{remark}
In what follows, the detailed proof is given for lattice potentials $V_N$,
however, in the fluid case the only difference is that  the
number of particles, interacting with a given one, is not preassigned.
For this reason, in the fluid case, the number of particles within the
interaction range of any other particle has to be replaced by its average.
After the end of Section \ref{conesempio}, more comments are given on this point.
\end{remark}

\subsubsection{Proof of $\sup_{N,\vb\in {I_{\vb}}}
\left | S_N({\vb}) \right | < \infty$}

This directly comes from the intensive character of $S_N$.
\qed


\medskip
\subsubsection{Proof of
$\sup_{N,\vb\in {I_{\vb}}} \left | \frac{\de S_N}{\de {\vb}}({\vb}) \right |
< \infty$}
\label{conesempio}
By definition of $S_N$ we have
    \bena
    \frac{\de
    S_N}{\de {\vb}}({\vb})
    = \frac{1}{N} \frac{\isnp}{\isn} \cdot \frac{d v}{d {\vb}} =
    \frac{\isnp}{\isn} \nonumber
    \eena

where $\isnp$ stands for the derivative of $\isn$ with respect to the potential
energy value $v=N \vb$.

The assumptions of our Main Theorem allow the use of the Federer-Laurence
theorem enunciated in Section \ref{proofLm2}
and of the derivation formula given therein, thus
    \bena
    \isnp = \int_{\Si_v}\ngV A \lpt \unv \rpt \frac{d \sigma}{\ngV}
        \dd ,
    \label{isnp}
    \eena
whence
    \bena
    \frac{\de
    S_N}{\de {\vb}}({\vb})
    = \frac{\isnp }{ \isn}
    = \left \langle \ngV A(1/\ngV) \right \rangle^{\mu c}_{N,v}
    \label{f1v_1}
    \eena
where $\left \langle~\right \rangle^{\mu c}_{N,v}$ stands for the
configurational microcanonical average performed on the equipotential
hypersurface of level $v$.

Let us proceed to show that this derivative is bounded by a term which is
independent of $N$.

To ease notations we define
    \bena
    \chi \equiv \unv \dd \label{cchi}
    \eena
so that Eq. (\ref{f1v_1}) now reads
    \bena
    \frac{\de
    S_N}{\de {\vb}}({\vb})
    = \left \langle \frac{1}{\chi} A(\chi) \right \rangle^{\mu c}_{N,v}~.
    \label{f1v_1p}
    \eena
It is
\bena
\frac{1}{\chi} A(\chi)= \frac{\Delta V}{\ngV^2}
  -2 \frac{\sum_{i,j=1}^N\partial^i V\partial^2_{ij} V \partial^j V}{\ngV^4}
\label{achischi}
\eena
and hence
\bena
\left \vert \frac{1}{\chi} A(\chi) \right \vert \leq
\frac{\mid \Delta V \mid}{ \ngV^2}
  +2 \frac{\mid \sum_{i,j=1}^N\partial^i V\partial^2_{ij} V \partial^j V \mid }{\ngV^4}~,
\nonumber
\eena
where $\partial_iV=\partial V/\partial q^i$, $q^i$ being the $i$-th coordinate
of configuration space ${\Bbb R}^N$.

In the absence of critical points of $V$ it is $\Vert\nabla V\Vert^2\geq C>0$, thus we
can apply Lemma \ref{comportement_ngV}, where $Y>0$ is required, to find
\bena
\left\vert \frac{\de S_N}{\de {\vb}}({\vb})\right\vert
&=&\left \vert \left \langle \frac{1 }{ \chi} A(\chi)
\right \rangle^{\mu c}_{N,v} \right \vert
\leq
\left \langle \left \vert \frac{1}{\chi} A(\chi) \right \vert
\right \rangle^{\mu c}_{N,v} \nonumber \\
&\leq&
\left \langle \frac{\mid \Delta V \mid}{\ngV^2} \right \rangle^{\mu c}_{N,v}
+2 \left \langle  \frac{\mid \sum_{i,j=1}^N\partial^i V\partial^2_{ij} V \partial^j V \mid
}{ \ngV^4} \right \rangle^{\mu c}_{N,v} \nonumber \\
& \leq&
\frac{\left \langle \mid \Delta V \mid \right \rangle^{\mu c}_{N,v}}
{\langle \ngV^2 \rangle^{\mu c}_{N,v}}
\ + O\left(\frac{1}{N}\right)
+2 \frac{\left \langle \sum_{i,j=1}^N \mid \partial^i V\partial^2_{ij} V \partial^j V \mid
\right \rangle^{\mu c}_{N,v}}{ \langle \ngV^4\rangle^{\mu c}_{N,v}}\
+ O\left(\frac{1}{N^2}\right)\ . \nonumber
\eena
Consider now the term $\langle \mid \Delta V \mid \rangle^{\mu c}_{N,v}$. As the potential $V$
is assumed smooth and bounded below, one has
\bena
\langle \mid \Delta V \mid \rangle^{\mu c}_{N,v}=
\left \langle \left \vert \sum_{i=1}^N \partial_{ii}^2 V \right \vert
\right \rangle^{\mu c}_{N,v}\ \leq
\sum_{i=1}^N \langle \mid \partial_{ii}^2 V \mid \rangle^{\mu c}_{N,v}
\ \leq N~\max_{i=1,..,N} \left \langle  \mid \partial_{ii}^2 V \mid
 \right \rangle^{\mu c}_{N,v}\ .
\nonumber
\eena
As a consequence of Remark \ref{fattorizza}, at large $N$
(when the fluctuations of the averages are vanishingly small)
$\max_{i=1,..,N}\langle \mid \partial_{ii}^2 V \mid \rangle^{\mu c}_{N,v}$
does not depend on $N$.
The same holds for
$\left \langle \mid \partial^i V\partial^2_{ij} V \partial^j V
\mid\right\rangle^{\mu c}_{N,v}$ and
$\max_{i=1,..,N} \left \langle \mid \partial^i V\partial^2_{ij} V \partial^j
V \mid \right \rangle^{\mu c}_{N,v}$.\\
We set
$m_1=\max_{i=1,..,N} \langle \mid \partial_{ii}^2 V \mid \rangle^{\mu c}_{N,v}$
and
$m_2=\max_{i,j=1,..,N} \left \langle \mid \partial^i V\partial^2_{ij} V
\partial^j V \mid \right \rangle^{\mu c}_{N,v}$

Let us now consider the terms $\langle \ngV^{2n} \rangle^{\mu c}_{N,v}$ for $n=1,2$.
One has
\bena
\langle \ngV^2 \rangle^{\mu c}_{N,v}=
\left\langle\sum_{i=1}^N ( \partial_i V )^2 \right\rangle^{\mu c}_{N,v}
\ = \sum_{i=1}^N \left\langle ( \partial_i V )^2\right\rangle^{\mu c}_{N,v}
\geq N~\min_{i=1,..,N}
\left \langle \left( \partial_i V \right)^2 \right \rangle^{\mu c}_{N,v}\ ,
\nonumber
\eena
\bena
\langle \ngV^4 \rangle^{\mu c}_{N,v}&=&
\left\langle \left[\sum_{i=1}^N ( \partial_i V )^2\right]^2 \right\rangle^{\mu c}_{N,v}
\ = \sum_{i,j=1}^N \left\langle ( \partial_i V )^2( \partial_j V )^2\right\rangle^{\mu c}_{N,v}
\nonumber\\
&\geq & N^2~\min_{i,j=1,..,N}
\left \langle \left( \partial_i V \right)^2 \left( \partial_j V \right)^2
\right \rangle^{\mu c}_{N,v}\ ,
\nonumber
\eena
By setting $c_1=\min_{i=1,..,N}\left \langle \left( \partial_i V \right)^2
\right \rangle^{\mu c}_{N,v}$
and $c_2=\min_{i,j=1,..,N}
\left \langle \left( \partial_i V \right)^2 \left( \partial_j V \right)^2
\right \rangle^{\mu c}_{N,v}$ we can finally write
\be
\left \vert \left \langle \frac{1}{\chi} A(\chi) \right \rangle^{\mu c}_{N,v}
\right \vert\leq
\frac{m_1}{c_1}+ O\left( \frac{1}{N} \right)\
+2 \frac{n_p~m_2}{c_2N}+ O \left( \frac{1}{N^2} \right)\
\label{maj_achischi}
\ee
where $n_p$ is the number of nearest neighbors. It is evident that in the limit
$N \to \infty$ the r.h.s. of the equation above tends to the finite constant $m_1/c_1$.

The upper bound thus obtained ensures that
$\sup_{N,\vb \in {I_{\vb}}} \left |\frac{\de S_N}{\de {\vb}}({\vb}) \right |<
\infty$.
\qed

\begin{remark}
Notice that, in the fluid case, the computation of quantities like
$\langle(\partial_iV)^2\rangle^{\mu c}_{N,v}$ or
$\langle\vert\partial_{ii}^2V\vert\rangle^{\mu c}_{N,v}$ involves
an a-priori unknown number of neighbors of the $i$-th particle
(we say that a particle is a neighbor of another
one if the distance between the two particles is smaller than the
interaction range of the potential). However, the requirement that $V$
is repulsive at short distance, so that clusters of an arbitrary number of
particles are forbidden, guarantees that each particle has a finite
average number of neighbors. Thus, averaging quantities like the above
mentioned ones yields $N$-independent values.

In order to extend to the fluid case the proofs of uniform boundedness
of the derivatives of the entropy (given throughout the present Section
\ref{parteB}), one has to interpret $n_p$ as the average number of neighbors
of a given particle.
\end{remark}

\begin{remark}
Notice that the above computations show that
\begin{eqnarray}
\lim_{N \too \infty} \left \langle \frac{A(\chi)}{\chi}
\right \rangle^{\mu c}_{N,v} ={\rm const} < \infty \nonumber
\end{eqnarray}
\label{achischi_o1}
\end{remark}
which follows from the boundedness of $\vert\langle A(\chi )/\chi\rangle\vert$.


\subsubsection{Proof of
$\sup_{N,\vb\in {I_{\vb}}} \left | \frac{\de^2 S_N}{\de {\vb}^2}({\vb})
\right | < \infty$}

The second derivative of $S_N$ can be rewritten in the form
    \bena
    \frac{\de^2 S_{N}}{\de {\vb}^2}({\vb})
    &=& N \cdot \lpq \frac{\Omega^{\prime\prime}(v,N)}{\isn} -
    \lpt \frac{\isnp}{\isn}
    \rpt^2 \rpq \dd
    \label{f2v_1}
    \eena
or, by using  the same notations as before,
    \bena
    \frac{\de^2 S_{N}}{\de {\vb}^2}({\vb}) =
      N  \left \{ \left \langle \frac{1}{\chi} A^2 \left
    ( {\chi}
    \right ) \right \rangle_{N,v}^{\mu c}\!\! \! \!\! \! - \left [
    \left \langle \frac{1}{\chi} A
    \left ( \chi \right ) \right
    \rangle_{N,v}^{\mu c} \right ]^2
    \right \} \dd
    \label{f2v_2}
    \eena
again we are going to show that an upper bound, independent of $N$, exists
also for this derivative.
In order to make notations compact, we define
\bena
\ps &\equiv& \frac{\nabla}{\ngV} \nonumber \\
for~any~ h_1, h_2,~\ps(h_1)~.~ \ps(h_2) &=& \sum_{i=1}^{N} \psi_i(h_1)
\psi_i(h_2)
\nonumber
\eena
whence simple algebra yields
    \bena
    \psvd \psc &=& \chi^2 M_1 - \chi^3 \dv \dd , \label{rel1} \\
    \underline{\psi}^2(V) &\equiv& \ps \lpt \dpsv \rpt = \frac{1}{\chi}
    \psvd \psc + \chi^2 \dv \label{rel2} \\
    \psi_i (\psvj) &=& \chi^2 \de^2_{ij} V -
     \chi^2 \psvj \sum_{k=1}^N\psvk \de^2_{ik} V  \label{rel3} \\
    \psci &=& - \chi^3 \sum_{j=1}^N\de^2_{ij}V \psvj  \label{rel4} \\
    \psi_i \lpt \psvj \rpt &=& \chi^2 \de^2_{ij}V - \chi^2
    \psvj \sum_{k=1}^N\psvk \de^2_{ik}V \label{rel5} \\
    \psi_i \lpt \de^2_{jr}V\rpt &=& \chi \de^3_{ijr}V \label{rel6} \\
    \psi_i \lpt \de^2_{jj}V\rpt &=&\chi \de^3_{ijj}V \label{rel7}
    \eena
where $M_1=\nabla (\nabla V / \|\nabla V \|) \equiv
-N\cdot (mean ~curvature~of~\Sigma_v)$. With these notations we have
    \bena
    A^2(\chi) &=& A\lpt A(\chi) \rpt
    = A \lpt \psvd \psc + \chi^3 \dv \rpt  \nonumber \\
    &=& \frac{1}{\chi} \lpt A(\chi) \rpt^2 + \chi \psvd \ps \lpt
    \frac{A(\chi)}{\chi} \rpt \dd
    \label{A2}
    \eena
and thus Eq. (\ref{f2v_2}) now reads
    \bena
    \left | \frac{\de^2 S_{N}}{\de {\vb}^2} (\vb) \right |
    &\leq & N \left |
    \left \langle
         \left [ \frac{A \left ( {\chi} \right )}{\chi} \right ]^2
    \right \rangle_{N,v}^{\mu c}
    - \left [ \left \langle
        \frac{A \left ( \chi \right )}{\chi}
    \right \rangle_{N,v}^{\mu c} \right ]^2 \right | \nonumber \\
    &+& N \left | \left \langle
        \psvd \ps \lpt \frac{A(\chi)}{\chi} \rpt
    \right \rangle_{N,v}^{\mu c} \right |~.
    \label{f2v_3}
    \eena
By using the relations (\ref{rel1})-(\ref{rel7}), the term
$\frac{1}{\chi} A \left ( {\chi} \right )$ is rewritten as
    \bena
    \frac{A(\chi)}{\chi} &=& \frac{1}{\chi} \ps \lpt \dpsv \chi \rpt
      =  \frac{2}{\chi} \psvd \psc + \chi^2 \dv \nonumber \\
    &=& 2 \chi M_1 - \chi^2 \dv \nonumber \\
    &=& \frac{\triangle V}{\ngV^2}~-~2~\frac{\sum_{i,j=1}^N\partial^i V\partial^2_{ij} V
    \partial^j V}{\ngV^4}~.
    \eena
Now we consider the following inequalities
    \bena
\left\vert \left \langle \frac{\sum_{i,j=1}^N\partial^i V \partial^2_{ij} V \partial^j V
    }{ \ngV^4} \right \rangle_{N,v}^{\mu c} \right\vert
 &\leq& \left \langle \frac{\left\vert\sum_{\langle i,j \rangle~;~i,j=1}^N
      \partial^i V\partial^2_{ij} V \partial^j V\right\vert
     }{\ngV^4} \right \rangle_{N,v}^{\mu c} \nonumber \\
    &\leq&\frac{\sum_{\langle i,j \rangle~;~i,j=1}^N
      \left \langle \vert\partial^i V\partial^2_{ij} V \partial^j V
     \vert\right \rangle_{N,v}^{\mu c} }{\left \langle\ngV^4\right \rangle_{N,v}^{\mu c} }
     + O\left(\frac{1}{N^2}\right)\nonumber \\
    &\leq& \frac{N~n_p~m_2}{c_2N^2}+ O\left(\frac{1}{N^2}\right)
     \label{majoration-N1}
    \eena
where $n_p$ is the number of nearest neighbours, and again

$m_2=\max_{i,j=1,..,N} \left \langle \mid \partial^i V\partial^2_{ij} V
\partial^j V \mid \right \rangle^{\mu c}_{N,v}$.

As $m_2$ keeps a finite value for $\lim_{N \to \infty}$,
the l.h.s. of equation
(\ref{majoration-N1}) vanishes in the $N\to\infty$ limit.

Thus, the larger $N$ the better the term
$\frac{1}{\chi} A \left ( {\chi} \right )$ is approximated by
$\xi = \sum_{i=1}^N \de^2_{ii} V / \ngV^2= \sum_{i=1}^N \xi_i$ where
$\xi_i=\de^2_{ii} V / \ngV^2$.
Here we resort to the Lemma \ref{mesure_ergodique} and
replace the microcanonical averages by ``time'' averages obtained along
an ergodic stochastic process. Each term $\xi_i$, for any $i$, can be then
considered as a stochastic process on the manifold $\Si_v$ with a
probability density $u_i(\xi_i)$.
In presence of short range potentials, as prescribed in the
hypotheses of our Main Theorem, and at large $N$, these processes are
independent.

By simply writing $\xi = \sum_{i=1}^N \xi_i = 1/N \sum_{i=1}^N N \xi_i$,
we are allowed to apply Lemma \ref{m234_borne} which tells us that the
the second moment $B_N'$ of the distribution of  $\xi$ is such that
$\lim_{N \to\infty} N~B_N' = c <\infty$.

The first term of the r.h.s. of (\ref{f2v_3}) is the second moment of
$\frac{1}{\chi} A \left ( {\chi} \right )$ multiplied by $N$, this term,
in the light of what we have just seen, remains finite in the $N \to \infty$
limit.

Then we consider the second term of the r.h.s. of equation (\ref{f2v_3}).
This can be computed with simple algebra through the relations
(\ref{rel1}-\ref{rel7}) to give
    \bena
    \psvd \ps \lpt \frac{A(\chi)}{\chi} \rpt
    &=& 8 \chi^4  \lpt \lps \psv ; \psv  \rps \rpt^2
    - 4 \chi^4 \lps \psv | \psv  \rps \nonumber \\
    &-& 2 \chi^4 \lps \psv ; \psv  \rps \triangle V
    + \chi^3 \sum_{i,j=1}^N\psvi \de^3_{ijj}V \nonumber \\
    &-& 2 \chi^3\sum_{i,j,k=1}^N \psvi \psvj \psvk \de^3_{ijk}V \dd
    \label{A2-terme2}
    \eena
where
    \bena
    \lps \psv ; \psv  \rps &\equiv&
    \frac{\sum_{i,j=1}^N\de_i V \de^2_{ij}V \de_j V }{\ngV^2}
    \label{partie1a} \\
    \lps \psv | \psv  \rps &\equiv&
    \frac{\sum_{i,j,k=1}^N\de_i V \de^2_{ij}V \de^2_{jk}V \de_k V }{\ngV^2}
    \label{partie2a}\\
    \psvi  \de^3_{ijj}V &\equiv& \frac{\de_i V \de^3_{ijj}V }{\ngV}
    \label{partie3a}\\
    \psvi \psvj \psvk \de^3_{ijk}V &\equiv&
    \frac{\de_i V  \de_j V \de_k V \de^3_{ijk}V }{\ngV^3}~.
    \label{partie4a}
    \eena
The same kind of computation developed for equations
(\ref{majoration-N1}) gives
    \bena
    N \left \langle
      \chi^4  \lpt \lps \psv ; \psv  \rps \rpt^2
    \right \rangle_{N,v}^{\mu c} &\leq&
        \frac{N^3 n_p^2 m_4}{c_4 N^4}
         +O \lpt \frac{1}{N^2} \rpt~~~
    \label{partie1b} \\
    N \left \langle
    \chi^4 \lps \psv | \psv  \rps
    \right \rangle_{N,v}^{\mu c} &\leq&
        \frac{N^2 n_p^2 m_5}{c_3 N^3}
         +O \lpt \frac{1}{N^2} \rpt~~~
    \label{partie2b}\\
    N \left \langle
      \chi^4 \lps \psv ; \psv  \rps \triangle V
    \right \rangle_{N,v}^{\mu c} &\leq&
        \frac{N^3 n_p m_6}{c_3 N^3}
         +O \lpt \frac{1}{N} \rpt~~~
    \label{partie1bp}\\
    N \left \langle
      \chi^3\sum_{i,j=1}^N \psvi \de^3_{ijj}V
    \right \rangle_{N,v}^{\mu c}  &\leq&
        \frac{N^2 n_p m_7}{ c_2 N^2}
         +O \lpt \frac{1}{N} \rpt~~~
     \label{partie3b}\\
    N \left \langle
      \chi^3\sum_{i,j,k=1}^N \psvi \psvj \psvk \de^3_{ijk}V
    \right \rangle_{N,v}^{\mu c}  &\leq&
        \frac{N^2 n_p^2 m_8}{c_3 N^3}
         +O \lpt \frac{1}{N^2} \rpt~~~
     \label{partie4b}
    \eena
where, resorting again to the argument of Remark \ref{fattorizza},
we have defined the following quantities independent of $N$
\[
m_4=\max_{i,j,k,l=1,N}\left\langle (\de_i V \de^2_{ij}V \de_j V)
(\de_k V \de^2_{kl}V \de_l V) \right \rangle_{N,v}^{\mu c}
\]
\[
m_5=\max_{i,j,k=1,N}\left\langle \de_i V \de^2_{ij}V \de^2_{jk}V \de_k V
 \right \rangle_{N,v}^{\mu c}
\]
\[
m_6=\max_{i,j,k=1,N}\left\langle (\de_i V \de^2_{ij}V \de_j V)
(\de^2_{kk}V )  \right \rangle_{N,v}^{\mu c}
\]
\[
m_7=\max_{i,j=1,N}\left\langle \de_i V \de^3_{ijj}V  \right \rangle_{N,v}^{\mu c}
\]
\[
m_8=\max_{i,j,k=1,N}\left\langle (\de_i V \de_jV \de_k V)
\de^3_{ijk}V   \right \rangle_{N,v}^{\mu c}
\]
and
\[
c_3=\min_{i_1,\dots,i_6=1,N}\left\langle (\de_{i_1} V)^2 (\de_{i_2} V)^2 \cdots
(\de_{i_6} V)^2\right \rangle_{N,v}^{\mu c}
\]
\[
c_4=\min_{i_1,\dots,i_8=1,N}\left\langle (\de_{i_1} V)^2 (\de_{i_2} V)^2 \cdots
(\de_{i_8} V)^2\right \rangle_{N,v}^{\mu c}
\]
so that the r.h.s. of Eqs. (\ref{partie1bp}) and (\ref{partie3b}) have finite limits
for $N\to\infty$, while the r.h.s. of (\ref{partie1b}), (\ref{partie2b}) and (\ref{partie4b})
vanish in the limit $N\to\infty$.

In conclusion, since the ensemble of terms entering equation (\ref{f2v_3}) is
bounded above, we have
$\sup_{N,\vb\in {I_{\vb}}} \left | \frac{\de^2 S_N}{\de {\vb}^2}({\vb})
\right | < \infty$.~\qed

\begin{remark}
Notice that the above computations show that
\begin{eqnarray}
\lim_{N \too \infty} N~\left\langle\psvd \ps \lpt \frac{A(\chi)}{\chi} \rpt
\right \rangle^{\mu c}_{N,v}= {\rm const}~
 <\infty ~.
~~ \nonumber
\end{eqnarray}
\label{p_o1sN}
\end{remark}

\subsubsection{Proof of
$\sup_{N,\vb\in {I_{\vb}}} \left | \frac{\de^3 S_N}{\de {\vb}^3}({\vb})
\right | < \infty$}

The third derivative of $S_N$ can be expressed as
    \bena
    &&\frac{\de^3 S_{N}}{\de {\vb}^3}({\vb})
    \nonumber \\
            &=& N^2 \left \{ \frac{\Omega^{\prime\prime\prime}(v,N)}{
    \Omega(v,N)}
    ~-~3 \frac{ \Omega^{\prime\prime}(v,N)
    \Omega^\prime (v,N)}{ (\Omega(v,N))^2}
    ~+~2~\left ( \frac{\Omega^\prime (v,N)
    }{ \Omega(v,N)} \right )^3 \right \} \nonumber
    \eena
or, by using Federer's operator $A$,
    \bena
    &&\frac{\de^3 S_{N}}{\de {\vb}^3}({\vb}) \\
    &=&\! N^2\! \left \{ \left \langle \frac{A^3(\chi)}{\chi}
    \right \rangle_{N,v}^{\mu c}\!\!\!\! \!\!-\!
    3 \! \left \langle \frac{A^2(\chi)}{\chi} \right \rangle_{N,v}^{\mu c}
    \left \langle \frac{A(\chi)}{\chi}
    \right \rangle_{N,v}^{\mu c}\!\!\!\! \!\!+\!
    2  \! \left ( \left \langle \frac{A(\chi)}{\chi}
     \right \rangle_{N,v}^{\mu c} \right )^3 \right \}\nonumber
     \label{ft3}
    \eena
where
    \bena
    \frac{A^3(\chi)}{\chi} &=& \left ( \frac{A(\chi)}{\chi} \right )^3 +
    3~\frac{A(\chi)}{\chi}~\psv \cdot \ps
    \left( \frac{A(\chi)}{\chi} \right ) \nonumber  \\
    &+&
    \psv \cdot \ps \left( \psv \cdot \ps \left( \frac{A(\chi)}{\chi}
     \right ) \right )\dd \label{a33} \\
     \frac{A^2(\chi)}{\chi} &=& \left ( \frac{A(\chi)}{\chi} \right )^2 +
    \psv \cdot \ps \left( \frac{A(\chi)}{\chi} \right ) \label{a23} \\
    \frac{A(\chi)}{\chi} &=& \frac{2}{\chi} \psv \cdot \psc+
    \frac{\triangle V}{\Vert \nabla V \Vert^2}~.  \label{a13}
    \eena
By substituting the expressions (\ref{a33})-(\ref{a13}) into the r.h.s. of
equation (\ref{ft3}), we get
    \bena
    &&\left | \frac{\de^3 S_{N}}{\de {\vb}^3}({\vb})
    \right | \nonumber \\
    &\leq& N^2 \left | \left \langle \psv \cdot \ps \left( \psv \cdot
    \ps \left( \frac{A(\chi)}{\chi} \right ) \right )
    \right \rangle_{N,v}^{\mu c} \right | \nonumber \\
    &+& 3 N^2 \left | \left \langle \frac{A(\chi)}{\chi}
    \psv \cdot \ps \left( \frac{A(\chi)}{\chi}
    \right ) \right \rangle_{N,v}^{\mu c}\!\! \!\!\!\!\!\!
    - \!\left \langle
    \frac{A(\chi)}{\chi} \right \rangle_{N,v}^{\mu c}\!\!\!
    \left \langle \psv \cdot \ps \left( \frac{A(\chi)}{\chi}
    \right ) \right \rangle_{N,v}^{\mu c} \right | \nonumber \\
    &+& N^2 \left | \left \langle  \lpt
    \left( \frac{A(\chi)}{\chi} \right ) -
    \left \langle \left( \frac{A(\chi)}{\chi} \right )
    \right \rangle_{N,v}^{\mu c}  \rpt^3
    \right \rangle_{N,v}^{\mu c} \right |~.
    \label{mostro}
    \eena
By explicitly expanding the first term of the r.h.s. of (\ref{mostro}) more
than $30$ terms are found. Nevertheless, these terms are similar or equal to
those already encountered above and, consequently, their $N$-dependence can
be similarly dominated as in the inequalities (\ref{partie1b}-\ref{partie4b}).

Consider now the second term of the r.h.s. of equation (\ref{mostro}).
If we put
\begin{eqnarray}
{\cal A} = \frac{A(\chi)}{\chi}~~~~~~~~~
{\cal P} = \psv \cdot \ps \left( \frac{A(\chi)}{\chi} \right ) \nonumber
\end{eqnarray}
using equations (\ref{achischi}) and (\ref{A2-terme2}) we can write
\begin{eqnarray}
{\cal A} = \sum_{i=1}^N a_i~~~~~~~~~{\cal P} = \sum_{j=1}^N p_j~. \nonumber
\end{eqnarray}
Then
\begin{eqnarray}
&&\left\langle \frac{A(\chi)}{\chi}~\psv
\cdot \ps \left( \frac{A(\chi)}{\chi} \right) \right\rangle_{N,v}^{\mu c}
- \left\langle \frac{A(\chi)}{\chi}
\right\rangle_{N,v}^{\mu c}\!\!\!
\left\langle \psv \cdot \ps \left( \frac{A(\chi)}{\chi}
\right) \right\rangle_{N,v}^{\mu c} \nonumber \\
&=& \langle {\cal A} {\cal P} \rangle_{N,v}^{\mu c}
- \langle {\cal A} \rangle_{N,v}^{\mu c}
\langle {\cal P} \rangle_{N,v}^{\mu c} \nonumber \\
&=& \sum_{i,j=1}^N \lpt \langle a_i p_j \rangle_{N,v}^{\mu c}
- \langle a_i \rangle_{N,v}^{\mu c}
\langle p_j \rangle_{N,v}^{\mu c} \rpt~.
\label{somme_aipj}
\end{eqnarray}
Let us consider the terms, in the last sum, for which $i$ and $j$ label sites
which are not nearest-neighbours\cite{nota2}.
The corresponding expressions of $a_i$ and $p_j$ have no common coordinate
variables. Thus, when computing microcanonical averages through ``time''
averages along the random Markov chains of Lemma \ref{mesure_ergodique},
we take advantage of the complete decorrelation of $a_i$ and $p_j$ so that
\begin{eqnarray}
for~any~i,j~s.t.
~0 \leq i,j \leq N,~~\rangle i,j\langle~ then~
\langle a_i p_j \rangle_{N,v}^{\mu c} - \langle a_i \rangle_{N,v}^{\mu c}
\langle p_j \rangle_{N,v}^{\mu c}=0 \nonumber
\end{eqnarray}
(where $\rangle i,j\langle$ stands for $i,j$ non nearest neighbours) which
simplifies equation (\ref{somme_aipj}) to
\begin{eqnarray}
 \langle {\cal A} {\cal P} \rangle_{N,v}^{\mu c}
- \langle {\cal A} \rangle_{N,v}^{\mu c}
\langle {\cal P} \rangle_{N,v}^{\mu c}
&=& \sum_{\langle i,j \rangle} \lpt \langle a_i p_j \rangle_{N,v}^{\mu c}
- \langle a_i \rangle_{N,v}^{\mu c}
\langle p_j \rangle_{N,v}^{\mu c} \rpt \nonumber \\
&\leq& N~n_p~\max_{\langle i,j \rangle} \lpt \langle a_i p_j
\rangle_{N,v}^{\mu c} - \langle a_i \rangle_{N,v}^{\mu c}
\langle p_j \rangle_{N,v}^{\mu c} \rpt~. \nonumber
\end{eqnarray}
Now, equations (\ref{maj_achischi}) and (\ref{partie1b}-\ref{partie4b})
imply
\begin{eqnarray}
for~any~i,j~s.t.~ 0 \leq i,j \leq N,~~\langle i,j\rangle~~
\lim_{N \too \infty} N^3~\langle a_i p_j \rangle_{N,v}^{\mu c}
  < \infty \nonumber
\end{eqnarray}
while equations (\ref{achischi}) and (\ref{A2-terme2}) imply
\begin{eqnarray}
for~any~i,j~s.t.~ 0 \leq i,j \leq N,~~\langle i,j\rangle~~~
\lim_{N \too \infty} N^3~\langle a_i \rangle_{N,v}^{\mu c}
\langle p_j \rangle_{N,v}^{\mu c}  < \infty~, \nonumber
\end{eqnarray}
where $\langle i,j\rangle$ stands for $i,j$ nearest neighbours.
Thus, the second term in the r.h.s. of equation (\ref{mostro}) is bounded
independently of $N$ in the limit $N\rightarrow\infty$. \\
The third term of the r.h.s. of equation (\ref{mostro}) is smaller than the
third moment of the stochastic variable $A(\chi)/ \chi$ (multiplied by
$N^2$). As we have already seen, we can rewrite
$A(\chi)/ \chi = (1/N) \sum_{i=1}^N N \partial_{ii}^2 V / \ngV^2$ to which
Lemma \ref{m234_borne} applies thus ensuring that the third moment $C_N'$
of the distribution of $A(\chi)/ \chi $ is such that
$\lim_{N \too \infty} N^2~C_N' = 0$.

Finally we are left with a finite upper bound of the l.h.s. of equation
(\ref{mostro}) in the $N\rightarrow\infty$ limit.~\qed

\begin{remark}
Notice that the computations above show that
\begin{eqnarray}
\lim_{N \too \infty} N^2~ \left\langle\psv \cdot \ps \left( \psv \cdot
\ps \left( \frac{A(\chi)}{\chi} \right ) \right )\right \rangle^{\mu c}_{N,v}
  = {\rm const} <\infty~.
~~ \nonumber
\end{eqnarray}
\label{w_o1sN2}
\end{remark}


\subsubsection{Proof of
$\sup_{N,\vb\in {I_{\vb}}} \left | \frac{\de^4 S_N}{\de {\vb}^4}({\vb})
\right | < \infty$}

The fourth derivative of $S_N(\vb)$ is given by the expression
\begin{eqnarray}
\frac{\de^4 S_N}{\de \vb^4}(\vb)&=& N^3 \left \{
\frac{ \Omega^{iv}(v,N)}{\Omega(v,N)}
- 4 \frac{ \Omega^{\prime\prime\prime}(v,N)~\Omega^\prime (v,N)
}{ \left ( \Omega(v,N) \right )^2}
- 3 \left ( \frac{ \Omega^{\prime\prime}(v,N)}{\Omega(v,N)} \right )^2
\right \} \nonumber \\
&+&N^3 \left \{
12 \frac{ \Omega^{\prime\prime}(v,N) \left (\Omega^\prime (v,N)
\right )^2}{\left ( \Omega(v,N) \right )^3 }
- 6 \left ( \frac{\Omega^\prime (v,N)}{ \Omega(v,N)}
\right )^4 \right \}\nonumber
\end{eqnarray}
Again we make use of the Federer operator $A$ to rewrite it as
\begin{eqnarray}
\frac{\de^4 S_N}{\de \vb^4}(\vb)&=& N^3 \left \{
\left \langle \frac{A^4(\chi)}{\chi} \right \rangle_{N,v}^{\mu c}
-4 \left \langle \frac{A^3(\chi)}{\chi} \right \rangle_{N,v}^{\mu c}
\left \langle \frac{A(\chi)}{\chi} \right \rangle_{N,v}^{\mu c} \right \}
\nonumber \\
&-& N^3 \left \{ 3 \left ( \left \langle \frac{A^2(\chi)}{\chi}
\right \rangle_{N,v}^{\mu c} \right )^2
- 12 \left \langle \frac{A^2(\chi)}{\chi}
\right \rangle_{N,v}^{\mu c}
\left ( \left \langle \frac{A(\chi)}{\chi} \right \rangle_{N,v}^{\mu c}
\right )^2 \right \} \nonumber \\
&-& 6 N^3 \left ( \left \langle \frac{A(\chi)}{\chi}
\right \rangle_{N,v}^{\mu c} \right )^4  \nonumber
\end{eqnarray}
where, after trivial algebra,
\begin{eqnarray}
\frac{A^4(\chi)}{\chi} &=& \left ( \frac{A(\chi)}{\chi} \right )^4
+6  \left ( \frac{A(\chi)}{\chi} \right )^2~
\psi(V)~.~\psi \left( \frac{A(\chi)}{\chi} \right ) \nonumber \\
&+&3~\left ( \psi(V)~.~\psi \left( \frac{A(\chi)}{\chi} \right ) \right )^2
+4~\frac{A(\chi)}{\chi}~\psi(V)~.~\psi \left( \psi(V)~.~
\psi \left( \frac{A(\chi)}{\chi} \right ) \right ) \nonumber \\
&+&\psi(V)~.~\psi \left [ \psi(V)~.~\psi \left( \psi(V)~.~
\psi \left( \frac{A(\chi)}{\chi} \right ) \right ) \right ]~.
\end{eqnarray}
To make the notations more compact we use
\begin{eqnarray}
{\cal A} = \frac{A(\chi)}{\chi}~~~~~~~~
{\cal P} = \psi(V)~.~\psi \left( \frac{A(\chi)}{\chi} \right ) \nonumber \\
{\cal W} = \psi(V)~.~\psi \left( \psi(V)~.~
\psi \left( \frac{A(\chi)}{\chi} \right ) \right ) \nonumber
\end{eqnarray}
so that, using again equations (\ref{a33}-\ref{a23}), we obtain
\begin{eqnarray}
\left | \frac{\de^4 S_N}{\de\vb^4}(\vb) \right | &\leq&
N^3 \left |\left \langle \psi(V)~.~\psi({\cal W})
\right \rangle_{N,v}^{\mu c} \right | \nonumber \\
&+&3 N^3 \left | \left \langle {\cal P}^2 \right \rangle_{N,v}^{\mu c}
- \left ( \left \langle {\cal P} \right \rangle_{N,v}^{\mu c}
\right )^2 \right | \nonumber \\
&+& 4 N^3 \left | \left \langle {\cal A} {\cal W} \right \rangle_{N,v}^{\mu c}
- \left \langle {\cal A} \right \rangle_{N,v}^{\mu c}
\left \langle {\cal W} \right \rangle_{N,v}^{\mu c} \right |
\label{d4-final} \\
&+& 6 N^3 \left | \left \langle \lpt {\cal A} - \left \langle {\cal A}
\right \rangle_{N,v}^{\mu c} \rpt^2 ~ \lpt {\cal P} - \left \langle {\cal P}
\right \rangle_{N,v}^{\mu c} \rpt \right \rangle_{N,v}^{\mu c} \right |
\nonumber \\
&+&N^3 \left | \left \langle \left ( {\cal A} - \left \langle {\cal A}
\right \rangle_{N,v}^{\mu c} \right )^4 \right \rangle_{N,v}^{\mu c}
- 3 \left (
\left \langle \left ( {\cal A} - \left \langle {\cal A}
\right \rangle_{N,v}^{\mu c} \right )^2 \right \rangle_{N,v}^{\mu c}
\right )^2 \right |~. \nonumber
\end{eqnarray}
Consider the first term of equation (\ref{d4-final}).
It is an iterative term already considered for the third derivative. This
term stems from the application of the operator $\psi(V)\cdot\psi(\cdot)$
to the term ${\cal W}$ which in its turn stems from the application of the
same operator to the term ${\cal P}$.
The effect of this operator is to lower the $N$ dependence of the function
upon which it is applied by a factor $N$ (what is simply due to the factor
$1/\ngV^2$). Deriving with respect to $\vb$ brings about a factor $N$ in
comparison to the derivation with respect to $v$, therefore the first term
of equation (\ref{d4-final}) is of the same order of
$N^2\ \langle {\cal W} \rangle^{\mu c}_{N,v}$ and
consequently, according to the Remark \ref{w_o1sN2}, it has a finite upper
bound independent of $N$ in the limit $N\rightarrow\infty$.\\

Consider now the second term of the r.h.s. of equation (\ref{d4-final}).
The Remark \ref{p_o1sN} ensures that $\lim_{N \too \infty} N~\langle{\cal P}
\rangle^{\mu c}_{N,v}< \infty$. Moreover, after Lemma \ref{m234_borne}
\begin{eqnarray}
\lim_{N \too \infty} N^3 \left ( \left \langle {\cal P}
- \langle {\cal P} \rangle^{\mu c}_{N,v} \right \rangle^{\mu c}_{N,v}
\right )^2 < \infty~. \nonumber \\
\end{eqnarray}
Consider now the third term of the r.h.s. of equation (\ref{d4-final}).
The Remarks \ref{achischi_o1} and \ref{w_o1sN2} entail
$\lim_{N \too \infty}\langle {\cal A}\rangle^{\mu c}_{N,v} < \infty$ and
$\lim_{N \too \infty} N^2~\langle{\cal W}\rangle^{\mu c}_{N,v} < \infty$.
Thus, after Lemma \ref{m234_borne}
\begin{eqnarray}
\lim_{N \too \infty} N^\frac{1}{2} \left ( \left \langle {\cal A}
- \langle {\cal A} \rangle^{\mu c}_{N,v} \right \rangle^{\mu c}_{N,v}
\right ) < \infty \nonumber \\
\lim_{N \too \infty} N^\frac{5}{2} \left ( \left \langle {\cal W}
- \langle {\cal W} \rangle^{\mu c}_{N,v} \right \rangle^{\mu c}_{N,v} \right )
 < \infty~, \nonumber
\end{eqnarray}
whence
\begin{eqnarray}
\lim_{N \too \infty} &N^3& \left | \left \langle {\cal A} {\cal W}
\right \rangle_{N,v}^{\mu c} - \left \langle {\cal A}
\right \rangle_{N,v}^{\mu c} \left \langle {\cal W}
\right \rangle_{N,v}^{\mu c} \right | \nonumber \\
&=&\lim_{N \too \infty} N^3  \left | \left \langle {\cal A}
- \langle {\cal A} \rangle_{N,v}^{\mu c}
\right \rangle_{N,v}^{\mu c} \right |~~\left | \left \langle {\cal W}
- \langle {\cal W} \rangle_{N,v}^{\mu c} \right \rangle_{N,v}^{\mu c} \right |
< \infty~. \nonumber\\
\end{eqnarray}
Consider now the fourth term of the r.h.s. of equation (\ref{d4-final}).
If we write
\begin{eqnarray}
{\cal A}=\frac{1}{N} \sum_{i=1}^N a_i~~~~~
{\cal P}=\frac{1}{N^2 } \sum_{i=1}^N p_i \nonumber
\end{eqnarray}
with $a_i$ and $p_i$ terms of order $1$, we have
\begin{eqnarray}
&& N^3 \left | \left \langle \lpt {\cal A} - \left \langle {\cal A}
\right \rangle_{N,v}^{\mu c} \rpt^2 ~ \lpt {\cal P} - \left \langle {\cal P}
\right \rangle_{N,v}^{\mu c} \rpt \right \rangle_{N,v}^{\mu c} \right |
\nonumber \\
&=& \frac{1}{N} \sum_{i,j,k=1}^N \left \langle
\lpt a_i - \left \langle a_i \right \rangle_{N,v}^{\mu c} \rpt ~
\lpt a_j - \left \langle a_j \right \rangle_{N,v}^{\mu c} \rpt ~
\lpt p_k - \left \langle p_k \right \rangle_{N,v}^{\mu c} \rpt
\right \rangle_{N,v}^{\mu c} \nonumber \\
&=& \frac{1}{N} \sum_{\rangle i, j, k \langle} \left \langle
\lpt a_i - \left \langle a_i \right \rangle_{N,v}^{\mu c} \rpt ~
\lpt a_j - \left \langle a_j \right \rangle_{N,v}^{\mu c} \rpt ~
\lpt p_k - \left \langle p_k \right \rangle_{N,v}^{\mu c} \rpt
\right \rangle_{N,v}^{\mu c} \nonumber \\
&+& \frac{1}{N} \sum_{\langle i,j,k \rangle} \left \langle
\lpt a_i - \left \langle a_i \right \rangle_{N,v}^{\mu c} \rpt ~
\lpt a_j - \left \langle a_j \right \rangle_{N,v}^{\mu c} \rpt ~
\lpt p_k - \left \langle p_k \right \rangle_{N,v}^{\mu c} \rpt
\right \rangle_{N,v}^{\mu c} \nonumber
\end{eqnarray}
where $\rangle i, j, k \langle$ means that at least two of the three
indexes refer to non nearest neighbours sites, whereas $\langle i,j,k \rangle$
means that the three indexes are nearest neighbours.
If $i,j,k$ are such that $\rangle i, j, k \langle$ then at least two
of the three terms $a_i$, $a_j$ and $p_k$ have no common configurational
variables.
The microcanonical averages are again estimated according to
Lemma \ref{mesure_ergodique} through a stochastic process on the
configurational coordinates. The random processes associated with
$a_i$, $a_j$ and $p_k$ are thus completely decorrelated and one has
\begin{eqnarray}
&&for~any~i,j,k,~s.t.~\rangle i, j, k \langle, \nonumber \\
&&\left \langle
\lpt a_i - \left \langle a_i \right \rangle_{N,v}^{\mu c} \rpt ~
\lpt a_j - \left \langle a_j \right \rangle_{N,v}^{\mu c} \rpt ~
\lpt p_k - \left \langle p_k \right \rangle_{N,v}^{\mu c} \rpt
\right \rangle_{N,v}^{\mu c} =0~. \nonumber
\end{eqnarray}
Now, if we consider $i,j,k$ such that $\langle i,j,k \rangle$, the three terms
$a_i$, $a_j$ and $p_k$ are certainly correlated but we notice that there
are only $Nn_p^2$ terms of this kind. Thus we have
\begin{eqnarray}
&& \frac{1}{N} \sum_{\langle i,j,k \rangle} \left \langle
\lpt a_i - \left \langle a_i \right \rangle_{N,v}^{\mu c} \rpt ~
\lpt a_j - \left \langle a_j \right \rangle_{N,v}^{\mu c} \rpt ~
\lpt p_k - \left \langle p_k \right \rangle_{N,v}^{\mu c} \rpt
\right \rangle_{N,v}^{\mu c} \nonumber \\
&\leq& n_c^2 \max_{\langle i,k \rangle} \left \{
 \lpt a_i - \left \langle a_i \right \rangle_{N,v}^{\mu c} \rpt~,~
\lpt p_k - \left \langle p_k \right \rangle_{N,v}^{\mu c} \rpt
\right \}~. \nonumber
\end{eqnarray}
Since the terms $a_i$ and $p_k$ are of order $1$, the largest term of the
preceding equation is independent of $N$, we have thus found the upper bound
of the fourth term of the r.h.s. of equation (\ref{d4-final}).\\

Finally, the last term of the r.h.s. of equation (\ref{d4-final}) is the
fourth cumulant of the stochastic variable  $A(\chi) /\chi$ (multiplied
by $N^3$).
As already seen above, we write
$A(\chi)/ \chi = 1/N \sum_{i=1}^N N \partial_{ii}^2 V / \ngV^2$
so that Lemma \ref{m234_borne} applies and ensures that the distribution
of $A(\chi)/ \chi $ has a fourth cumulant $K_N'$ such that
$\lim_{N \too \infty} N^3~K_N' = 0$.

The ensemble of the upper bounds thus obtained yields the final desired result.
~\qed

\section{Final remarks}
To conclude this first paper, some comments are in order.

\begin{remark}[Domain of physical applications]
Notice that the requirement of  standard, stable, confining and short-range
potentials $V_N$ applies to a broad class of physically relevant
models. In fact, the interatomic and intermolecular interaction potentials
(like Lennard-Jones, Morse, van der Waals potentials)
which are typically encountered in condensed matter theory, as well as
classical spin potentials, fulfil these requirements.
\end{remark}

\begin{remark}[Sufficiency conditions]
\label{sufficiency} Notice that the converse of our Main Theorem
is not true, in other words there is not a one-to-one
correspondence between any topology change of the energy level sets
and phase transitions. In fact, there are systems, like the Fermi-Pasta-Ulam
model described by $V_N(q)= \sum_{i=1}^N\frac{1}{2}(q_{i+1} - q_i)^2+
\frac{\lambda}{4}(q_{i+1} - q_i)^4$ which, for fixed end points, has no
critical
points and no phase transitions, whereas, for example, a one
dimensional lattice of classical spins (or of coupled rotators)
described by the potential function $V_N(q)= \sum_{i=1}^N [1 -
\cos (q_{i+1} - q_i)]$ has many critical points \cite{xymf} so
that both families $\{\Sigma_v\}_{v\in{\Bbb R}}$ and
 $\{ M_v\}_{v\in{\Bbb R}}$ undergo many topology changes,
but, since no phase transition is associated with this potential, none of
these topology changes corresponds to a phase transition. Note that this is
not a counter
example of our Main Theorem (which would require to
find a system undergoing
a phase transition in the absence of topology changes and within the domain of
validity of the Theorem), it just tells us that  the loss of
diffeomorphicity of the $\{\Sigma_v\}_{v\in{\Bbb R}}$ and, equivalently,
of the $\{ M_v\}_{v\in{\Bbb R}}$ at some $v_c$,
is a {\it necessary} but {\it not sufficient} condition for the
occurrence of a phase transition.
\end{remark}

\begin{remark}[Relevance of topology changes for phase transitions]
\label{relevance}
In order to  prove that our Theorem is relevant to statistical mechanics,
and in particular in order to really link the phenomenon of phase transitions
to a
topology change of the configuration space submanifolds $M_v$, in paper II
we work out an analytic relation between configurational entropy $S(v)$ and the
Morse indexes of the submanifolds $M_v$. Such a relation is formulated within
another Theorem (enunciated also in the Introduction of the present paper)
which unveils why the differentiability class of $S(v)$, in the $N\to\infty$
limit, can be lowered from ${\cal C}^\infty$ to ${\cal C}^2$ or to ${\cal C}^1$
only by a suitable energy change of the Morse indexes (hence of topology
change).
Loosely speaking, in the context of our topological approach, the Theorem
proved in paper II plays an analogous role to that played by the Lee-Yang
circle Theorem \cite{LYthm} within the context of the Yang-Lee theory of
phase transitions.
\end{remark}

\section{Acknowledgments}
The authors wish to thank A. Abbondandolo, H. van Beijeren, L. Casetti,
C. Liverani, A. Moro, P. Picco for comments and suggestions.
A particularly warm acknowledgment is addressed to G. Vezzosi for his
continuous interest in our work and for many helpful discussions and
suggestions.




\begin{thebibliography}{99}

\bibitem{Pettini} M.\ Pettini, Phys.\ Rev.\ E {\bf 47}, (1993) 828.

\bibitem{pre96} L.\ Casetti, C.\ Clementi, and M.\ Pettini,
Phys.\ Rev.\ E {\bf 54}, (1996) 5969.

\bibitem{cccp} L.\ Caiani, L.\ Casetti, C.\ Clementi, and M.\ Pettini,
Phys.\ Rev.\ Lett.\ {\bf 79}, (1997) 4361.

\bibitem{pre98} L.\ Caiani, L.\ Casetti, C.\ Clementi, G.\ Pettini,
M.\ Pettini, and R.\ Gatto, Phys.\ Rev.\ E {\bf 57}, (1998) 3886.

\bibitem{jpa98} L.\ Caiani, L.\ Casetti, and M.\ Pettini,
J.\ Phys.\ A: Math.\ Gen.\ {\bf 31}, (1998) 3357.

\bibitem{CSCP} M. Cerruti-Sola, C. Clementi and  M. Pettini,
Phys.\ Rev.\ E {\bf 61}, (2000) 5171.

\bibitem{Firpo} M.-C.\ Firpo, Phys.\ Rev.\ E {\bf 57}, (1998) 6599.

\bibitem{top1} R.\ Franzosi, L.\ Casetti, L.\ Spinelli, and M.\ Pettini,
Phys.\ Rev.\ E {\bf 60}, (1999) R5009.

\bibitem{top3} L.\ Casetti, E.\ G.\ D.\ Cohen, and M.\ Pettini,
Phys.\ Rev.\ Lett.\ {\bf 82}, (1999) 4160.

\bibitem{top2} R. Franzosi, M. Pettini, and L.Spinelli,
               Phys. Rev. Lett. \textbf{84}, (2000) 2774.

\bibitem{xymf} L. Casetti, M. Pettini, and E.G.D. Cohen,
               J. Stat. Phys. \textbf{111}, (2003) 1091.

\bibitem{ptrig} L. Angelani, L. Casetti, M. Pettini, G. Ruocco, and F. Zamponi,
               Europhys. Lett.\textbf{62}, (2003) 775; Phys.\ Rev.\ E
               {\bf 71}, (2005) 036152.
\bibitem{physrep} L. Casetti, M. Pettini, and E.G.D. Cohen,
               Phys. Rep. \textbf{337}, (2000) 237-341.

\bibitem{pirl} R. Franzosi, and M. Pettini,
               Phys. Rev. Lett. \textbf{92}, (2004) 060601.

\bibitem{federer} H. Federer, \textit{Geometric Measure Theory}, (Springer,
                 New York 1969), p. 249.

\bibitem{ruelle} D. Ruelle, \textit{Statistical Mechanics. Rigorous results},
(Benjamin, Reading, 1969).

\bibitem{palais} R.S. Palais and C. Terng, \textit{Critical Point Theory and
                 Submanifold Geometry}, (Springer, New York 1988).

\bibitem{hirsch} M.W. Hirsch, \textit{Differential Topology}, (Springer,
                 New York 1976).

\bibitem{schwartz} L. Schwartz, \textit{Analyse. Topologie G\'en\'erale et
                  Analyse Fonctionelle}, (Hermann, Paris, 1970), Deuxi\`eme
                  Partie, p. 310.
\bibitem{milnor} J. Milnor, \textit{Morse Theory}, (Princeton University Press, Princeton, 1973).
\bibitem{bott} R. Bott and J. Mather, {\it Topics in Topology and
Differential Geometry}, in {\sl Battelle Rencontres},
Eds. C.M. De Witt and J.A. Wheeler, p.460.

\bibitem{laurence} P. Laurence, ZAMP \textbf{40}, (1989) 258.

\bibitem{thorpe} J.A. Thorpe, \textit{Elementary Topics in Differential
                 Geometry}, (Springer, New York, 1979), p. 150.


\bibitem{mcmc} P. Br\'emaud, \textit{Markov Chains},
                 (Springer-Verlag, New York 2001), chapter 7.

\bibitem{khinchin} A.I. Khinchin, \textit{Mathematical Foundations of
Statistical Mechanics}, (Dover Publications, Inc., New York 1949).

\bibitem{nota1} As at any finite $N$ all these functions are
${\cal C}^\infty$, the {\it supremum} always exists for finite $N$.

\bibitem{nota2} For simplicity we are here assuming
that the configurational coordinates belong to a lattice, but such a
restriction is not necessary. If our potential describes a fluid, replace
``nearest-neighbours'' with ``within the interaction range''.

\bibitem{LYthm} T.D. Lee and C.N. Yang, Phys. Rev. \textbf{87}, (1952) 410.


\end{thebibliography}
\end{document}